\documentclass[iicol,pdflatex,sn-mathphys-num]{sn-jnl}

\usepackage{graphicx}%
\usepackage{multirow}%
\usepackage{amsmath,amssymb,amsfonts}%
\usepackage{amsthm}%
\usepackage{mathrsfs}%
\usepackage[title]{appendix}%
\usepackage{xcolor}%
\usepackage{textcomp}%
\usepackage{manyfoot}%
\usepackage{booktabs}%
\usepackage{algorithm}%
\usepackage{algorithmicx}%
\usepackage{algpseudocode}%
\usepackage{listings}%
\usepackage{caption} 
\usepackage{placeins}
\usepackage{dcolumn}
\usepackage{bm}
\usepackage{hyperref}
\usepackage{mathtools}
\usepackage{subcaption}
\usepackage{array}
\usepackage{tabularx}
\usepackage[autostyle]{csquotes}
\MakeOuterQuote{"}
\usepackage{siunitx}
\DeclareSIUnit \h {\ensuremath{\mathit{h}}}
\DeclareSIUnit \parsec {pc}
\usepackage{afterpage}
\makeatletter

\DeclareRobustCommand{\diff}{%
  \@ifnextchar^{\DIfF}{\DIfF^{}}%
}
\newcommand{\DIfF}{}
\renewcommand{\DIfF}[1][]{%
  \mathop{\mathrm{\mathstrut d}}\nolimits^{#1}\gobblespace
}
\newcommand{\gobblespace}{%
  \futurelet\diffarg\opspace
}
\newcommand{\opspace}{%
  \let\DiffSpace\!%
  \ifx\diffarg(%
    \let\DiffSpace\relax
  \else
    \ifx\diffarg[%
      \let\DiffSpace\relax
    \else
      \ifx\diffarg\{%
        \let\DiffSpace\relax
      \fi
    \fi
  \fi
  \DiffSpace
}
\DeclareRobustCommand{\deriv}[3][]{%
  \frac{\diff^{#1} #2}{\diff #3^{#1}}%
}
\DeclareRobustCommand{\pderiv}[3][]{%
  \frac{\partial^{#1} #2}{\partial #3^{#1}}%
}

\newcommand{\vect}[1]{\boldsymbol{#1}}

\begin{document}

\title[Late-Time LSS Growth in Quasi-Coasting Models]{Testing Quasi-Linear Coasting Cosmologies with Late-Time Large-Scale Structure Growth}

\author*[1]{\fnm{D\'avid A.} \sur{K\"odm\"on}}\email{davidkodmon@student.elte.hu}

\author[1,2]{\fnm{Peter} \sur{Raffai}}\email{peter.raffai@ttk.elte.hu}

\affil*[1]{\orgdiv{Department of Atomic Physics}, \orgname{Institute of Physics and Astronomy,  ELTE E\"otv\"os Lor\'and University}, \orgaddress{\street{Pázmány Péter stny. 1/A}, \city{Budapest}, \postcode{1117}, \country{Hungary}}}

\affil[2]{\orgname{HUN-REN–ELTE Extragalactic Astrophysics Research Group}, \orgaddress{\street{Pázmány Péter stny. 1/A}, \city{Budapest}, \postcode{1117}, \country{Hungary}}}

\abstract{We derive analytical expressions for the growth factor, $D(z)$, and density-weighted growth rate, $f\sigma_8(z)$, for cosmologies in which $a\propto t$ at late times. We fit the resulting $f\sigma_8(z)$ predictions to redshift-space-distortion measurements in the range $z<2$ using the \texttt{dynesty} implementation of nested sampling. Three coasting models, with curvature parameters ${k=\{-1,0,+1\}}$ in $H_{0}^{2}c^{-2}$ units, and a flat $\Lambda$CDM model are tested. We evaluate each model's consistency with the data using the Anderson--Darling test for normality applied to the uncertainty-normalised residuals, supplemented by posterior predictive checks. For the coasting models, we obtain ${\Omega_\mathrm{m,0}=\{0.206^{+0.073}_{-0.061},\,0.297^{+0.085}_{-0.073},\,0.412^{+0.097}_{-0.086}\}}$ and ${\sigma_{8}(z=0)=\{1.071^{+0.213}_{-0.151},\,0.867^{+0.128}_{-0.097},\,0.725^{+0.080}_{-0.065}\}}$, respectively. For $\Lambda$CDM, we obtain $\Omega_\mathrm{m,0}=0.286^{+0.053}_{-0.047}$ and $\sigma_{8}(z=0)=0.764^{+0.039}_{-0.035}$. All models are consistent with the data, although $\Lambda$CDM is favoured over the coasting models, with log Bayes factors ${\log_{10}{\mathcal{B}}=\{1.79,\,1.55,\,1.42\}}$. This preference is robust against alternative priors and prior parametrisations, but weakens when heavier-tailed likelihoods are adopted. Predictive performance is assessed using the expected log predictive density computed by leave-one-out cross-validation. The $\Lambda$CDM model has the largest predictive performance, but its advantage is statistically significant only relative to the ${k=\{-1,0\}}$ coasting models.}

\maketitle

\section{Introduction} 
\label{sec:intro}
The concordance model of cosmology, commonly referred to as the Lambda Cold Dark Matter ($\Lambda$CDM) model,
has proven to be a highly robust theoretical framework for understanding the evolutionary history of the universe~\citep{cit:Bull}. In this model, the mechanisms governing the formation of large-scale structure (LSS)~\citep{cit:Bull}, and the cosmic microwave background (CMB) are well defined~\citep{cit:Turner}. Furthermore, the $\Lambda$CDM model provides an excellent fit to a variety of different cosmological observations, such as the \textit{Planck} CMB temperature (TT), polarisation (EE), and cross-power (TE) spectra~\citep{cit:Efstathiou}, as well as data from galaxy redshift surveys characterising LSS and the expansion history of the universe inferred from type Ia supernovae~\citep{cit:Turner}, to name a few. However, it is precisely these cosmological probes which pose empirical challenges to the model's validity, since the values of certain cosmological parameters inferred from these probes differ from one another~\citep{cit:cosmoverse,cit:Verde2024}. For example, the value of the Hubble constant $H_0$ determined via the distance ladder approach and the one inferred from CMB fluctuations differs by more than $5\sigma$~\citep{cit:Riess}; this has become known as the Hubble tension. Until recently, a tension between the value of the weighted amplitude of matter fluctuations, $S_8$, as determined from the CMB and from late-time probes, such as weak lensing, redshift-space distortions, and cluster counts was also identified as significant discrepancy between early- and late-time inference; this is known as the growth tension~\citep{cit:Perivolaropoulos}. 

Recent work on the problem has suggested that this tension is both probe- and survey-dependent~\citep{cit:Pantos2026}. In particular, while DES Year 6 continues to favour a relatively low value of $S_8$~\citep{cit:Abbott2026}, the final KiDS-Legacy cosmic-shear analysis finds a value of $S_8$ consistent with \textit{Planck}, with the shift relative to previous KiDS analyses attributed primarily to improved redshift calibration, increased survey area, and improved image reduction~\citep{cit:Wright2025}. Additionally, CMB-lensing measurements from ACT find structure-growth constraints consistent with $\Lambda$CDM expectations inferred from the primary CMB~\citep{cit:Madhavacheril2024}. These results suggest that the growth tension is not presently a single coherent discrepancy in the same sense as the Hubble tension. While observational and modelling systematics may play an important role, no single systematic effect has yet been shown to account for all low-$S_8$ measurements, and new physics has not been decisively excluded~\citep{cit:Pantos2026}.

The $\Lambda$CDM model also faces theoretical difficulties, such as its reliance on yet poorly understood cosmic constituents, such as dark matter and dark energy, which currently lie beyond the scope of the Standard Model of particle physics~\citep{cit:Turner}, and numerous other problems, such as the cosmic coincidence problem, as well as the cosmic age, synchronicity, horizon, and flatness problems~\citep{cit:Casado}. While the inflationary period addresses the horizon and flatness problems~\citep{cit:Guth}, the others remain unresolved. Given these theoretical and empirical challenges, exploring alternative cosmological models appears motivated.

The present study examines a family of models referred to as coasting cosmologies. The defining feature of these models is that scale factor, $a(t)$, is a linear function of cosmic time ($a\propto t$), although its realisation varies from model to model~\citep{cit:Casado}. Coasting cosmologies can be subdivided into strictly linear models, where $a \propto t$ holds at all times, and quasi-linear models, where the universe transitions from a $\Lambda$CDM model to a $a \propto t$ expansion sometime after recombination (for a review, see~\citet{cit:Casado}). Coasting models are also diversified by having different spacetime geometries. For example, the $R_\mathrm{h}=c\,t$ model~\citep{cit:Melia-Shevchuk, cit:Melia-book}, which is strictly linear, possesses a flat geometry (with curvature parameter $k=0$), while the eternal coasting model~\citep{cit:John-Joseph-2000} also admits spherical ($k>0$) or hyperbolic geometries ($k<0$). Strictly linear coasting cosmologies naturally resolve the horizon problem because the particle horizon diverges for $a \propto t$, and they reduce the synchronicity problem to a trivial consequence of linear expansion, as $a \propto t \Rightarrow H(t)t = 1$. Furthermore, strictly linear coasting cosmologies are also free from the flatness problem~\citep{cit:Casado}. However, strictly linear models also face various theoretical challenges. They cannot easily account for the observed abundances of light elements, while also lacking a robust model for the origin of the CMB's anisotropies and properties~(see~\citet{cit:Raffai} and references therein). In contrast, quasi-linear coasting models do not suffer from these problems, since they are essentially $\Lambda$CDM models prior to recombination, while also resolving the cosmic coincidence and fine-tuning problems~\citep{cit:Casado}. Coasting cosmologies are also demonstrably consistent with a variety of empirical data (for an extensive list, see Table 2 in~\citep{cit:Melia-RhBetter}), and are even slightly favoured over the $\Lambda$CDM model by late-time probes of cosmic expansion, such as type Ia supernovae, quasars, and cosmic chronometers~\citep{cit:Raffai2}. Hence, it appears natural to test quasi-linear cosmologies against other late-time probes as well.

While many aspects of specific coasting cosmologies have been extensively studied, the question of the growth of LSS has received less attention. The time evolution of LSS growth is expressed by the growth factor, $D(z)$ (see Section \ref{sec:Derivation}). An expression for $D(z)$ for a quasi-linear coasting universe in which linear expansion is realised through the dominance of an exotic component called K-matter, which has an equation of state parameter $w_\mathrm{K}=-1/3$, was derived by~\citet{cit:Kolb}, while \citet{cit:Rhct-Structure} performed a similar derivation in the context of the $R_\mathrm{h}=c\,t$ model. Each of these derivations relied on their specific realisations of linear expansion to motivate polynomial \textit{ans\"{a}tze} when solving the relevant differential equation for $D(z)$, yielding similar but distinct results. For the purposes of this work, we will derive an expression for $D(z)$ without making use of any property of a coasting universe other than the $a(t)\propto t$ functional form of its scale factor within the redshift range covered by empirical data.

To test the consistency of a theoretical $D(z)$ with observed LSS growth, it is expedient to use empirical data from redshift-space distortion~\citep{cit:Hamilton} measurements, expressed as the density-weighted growth rate, ${f\sigma_8(z)\equiv f(z)\sigma_8(z)}$, where $\sigma_8(z)$ is the amplitude of mass fluctuations in a sphere with a comoving radius of \SI{8}{\per\h\mega\parsec} (where $h\equiv H_0/\SI{100}{\kilo\meter\per\mega\parsec}$), and $f(z)$ is the growth rate~\citep{cit:Huterer}. Both $\sigma_8$ and $f$ are directly linked to $D$, since ${\sigma_8(z)\equiv\sigma_{8}(z=0)D(z)}$ and ${f(a)\equiv\diff{\ln{D}}/\diff{\ln{a}}}$~\citep{cit:Huterer}, which makes them ideal probes. However, empirical determinations of $\sigma_8$ and $f$ rely on measurements of galaxy clustering and peculiar velocities, and as such, they reflect the matter overdensity in galaxies, rather than the total matter overdensity~\citep{cit:Huterer}. The relationship between the two overdensities can be expressed using the galaxy bias factor, $b$, which is both scale and redshift dependent~\citep{cit:Huterer}. The empirical values of $\sigma_8$ and $f$ obtained using data from galaxies ($\sigma_{8,\mathrm{g}}$ and $f_\mathrm{g}$) are related to the true values as ${\sigma_{8,\mathrm{g}}=b\cdot\sigma_8}$ and ${f_\mathrm{g}=f/b}$~\citep{cit:Turner_RJ}. Obtaining $\sigma_8$ from $\sigma_{8,\mathrm{g}}$ or $f$ from $f_\mathrm{g}$ requires knowledge of the scale-dependence of $b$ (which usually has to be determined directly from the given galaxy sample), as well as \textit{a priori} information on the redshift-dependence of $b$, since the redshift-dependencies of $b$ and $D$ are degenerate~\citep{cit:Huterer}. However, ${f_\mathrm{g}\sigma_{8,\mathrm{g}}}$ is free from galaxy bias, and ${f_\mathrm{g}\sigma_{8,\mathrm{g}}=f\sigma_8}$, making it a preferable observable over both $\sigma_8$ and $f$.
A consistency test of the $R_\mathrm{h}=ct$ model using $f\sigma_8$ data published between 2011 and 2015 was performed by~\citet{cit:Rhct-Structure}. As there are now $f\sigma_8$ values for over forty distinct redshifts in the $z\in (0, 2)$ range, we will test our expression's consistency with $f\sigma_8$ data using a larger dataset, collated by~\citet{cit:Perivolaropoulos-Skara}.

By fitting our expression for $f\sigma_8(z)$ to the data, we can constrain the present values of the density parameter of matter, $\Omega_\mathrm{m,0}$, and the amplitude of mass fluctuations, $\sigma_8(z=0)$. Using these, we can constrain the weighted amplitude of matter fluctuations, which is defined as~\citep{cit:Huterer}
\begin{equation}
S_8\equiv\sigma_8(z=0)\sqrt{\frac{\Omega_\mathrm{m,0}}{0.3}}.
\end{equation}
Obtaining a value for $S_8$ in coasting cosmologies is necessary for us examining whether coasting models are free of the growth tension. Lastly, we will also perform this analysis for a $\Lambda$CDM model with \textit{Planck} 2018 "TT,TE,EE+lowE+lensing" (which we hereafter shorten to "baseline") best-fit parameters~\citep{cit:planck2018}, enabling us to determine whether the data exhibit a preference for either model.

Our treatment of LSS growth in coasting cosmologies will, in principle, be limited to quasi-linear models. Strictly linear models face significant challenges in explaining the shape of the linear matter power spectrum, $P(k)$. Notably, $P(k)$ is not monotonic with $k$ and, based on CMB observations, exhibits a maximum at the characteristic length scale corresponding to matter-radiation equality, $k_\mathrm{eq}$ (see \citet{cit:Planck-power_spectrum} for a visual representation). In the $\Lambda$CDM model, this extremum arises because perturbations of length scales exceeding that of the particle horizon (super-horizon perturbations) evolve at a different rate than those of length scales lying inside the horizon (sub-horizon perturbations) during the radiation-dominated epoch, whereas both super- and sub-horizon perturbations grow at the same rate during the matter-dominated era~\citep{cit:Padmanabhan}. The result is that the scale-independent primordial power spectrum produced by inflation is modified in a scale-dependent manner. This explanation does not apply to a strictly linear coasting universe, since it experiences uniform expansion throughout its history (without an inflationary period to seed the initial fluctuations) and lacks horizon crossing due to the absence of a particle horizon. Moreover, the peak in the matter power spectrum is identified based on analysing CMB anisotropies within the context of the concordance model. Therefore, it is possible that we would not observe a peak if the CMB fluctuation spectrum were constructed using a strictly linear model. 

\citet{cit:Yennapureddy2021} proposed an alternative mechanism for producing a turnover in $P(k)$ which relies on the scale-dependent evolution of perturbations in the ${R_\mathrm{h}=ct}$ model. In their treatment, the turnover is attributed to a scale-dependent balance between the decay of the perturbed gravitational potential and the growth timescale of the density modes. Shorter-wavelength modes experience more rapid potential decay, while longer-wavelength modes evolve on longer dynamical timescales and therefore grow more slowly. An intermediate scale can thus acquire the largest power, producing a turnover without invoking horizon exit and re-entry~\citep{cit:Yennapureddy2021}. However, beyond the ${a(t)\propto t}$ requirement, the mechanism relies on the ${R_\mathrm{h}=ct}$-specific condition ${\rho+3p/c^{2}=0}$, the assumed perturbability of the dark-energy component, and an interaction between dark energy and matter used to maintain the required cosmic evolution. These preclude its use as a general mechanism for all strictly linear coasting cosmologies. 

In contrast to strictly linear models, the formation of the CMB and its important features, as well as those of the matter power spectrum can be explained within quasi-linear coasting models by the same mechanisms as in the concordance model, provided that the transition to coasting expansion occurs after recombination. 
As our treatment of quasi-linear cosmologies aims to be as general as possible, we will perform consistency tests for coasting models with hyperbolic, flat, and spherical geometries, with curvature parameters ${k=\{-1, 0, +1\}}$ in $H^2_0c^{-2}$ units. Given that the available data cover a redshift range of $z \in (0\,,2)$, our results should be interpreted as tests of any cosmology undergoing approximately linear expansion within this interval, independently of the detailed expansion history at higher redshift, except insofar as that earlier history fixes the initial conditions of the low-redshift observables.

This paper is organised as follows. In Section~\ref{sec:Derivation}, we derive a closed form expression for $D(z)$ and $f\sigma_8(z)$ for coasting cosmologies in a manner consistent with the fluid treatment outlined by \citet{cit:Huterer-book}. In Section~\ref{sec:Data_and_Analysis}, we describe the method used to fit our formula to $f\sigma_8(z)$ data collated by \citet{cit:Perivolaropoulos-Skara}. We then explain how this procedure is adapted to the $\Lambda$CDM model, as well as the method used to assess the models' consistency with the data. In Section~\ref{sec:Results}, we discuss the results of the fit and the aforementioned tests. Finally, in Section~\ref{sec:Conclusions}, we summarise our conclusions and outline possibilities for expanding upon the analyses performed herein.

\section{Growth of Large Scale Structure in Coasting Cosmologies}
\label{sec:Derivation}
The evolution of LSS is captured by the density contrast function, $\delta$, defined as 
\begin{equation}
\delta(\vect{r},\,t)=\frac{\delta\rho(\vect{r},\,t)}{\overline{\rho}(t)},
\label{eq:def-delta}
\end{equation}
where $\overline{\rho}$ is the matter density of the homogeneous background, and $\delta\rho$ is the local deviation, representing an over- ($\delta\rho>0$) or underdensity ($\delta\rho<0$) relative to the background; their parameters are proper position, $\vect{r}$, and cosmic time, $t$. It is often expedient to work with $\delta$ in Fourier space. The differential equation governing the evolution of the various Fourier modes of $\delta$ is the following~\citep{cit:Huterer-book}

\begin{equation}
\begin{split}
\pderiv[2]{\delta_{\vect{k}}}{t}+2\,\frac{\dot{a}}{a}\pderiv{\delta_{\vect{k}}}{t}&=\left(4\pi G\overline{\rho}-\frac{c_\mathrm{s}^2|\vect{k}|^2}{a^2}\right) \delta_{\vect{k}}\\& -\frac{2}{3}\frac{\overline{T}}{a^2}S_{\vect{k}},
\end{split}
\label{eq:mastereq-K}
\end{equation}
where $G$ is the gravitational constant, $\vect{k}$ is the comoving wavevector associated with the given Fourier mode, $c_\mathrm{s}$ is the adiabatic sound speed, $\overline{T}$ is the temperature of the background, and $S_{\vect{k}}$ is the corresponding Fourier mode of the specific entropy, $S$. Since we are working with quasi-linear coasting cosmologies, we can use the fact that inflation typically predicts isentropic initial conditions to set $S_{\vect{k}}=0$~\citep{cit:Huterer-book}. Moreover, since the Jeans length, which is defined as ${\lambda_\mathrm{J} \equiv c_\mathrm{s}\sqrt{\pi/(G\overline{\rho})}}$, corresponds to a mass scale on the order of $10^5$ solar masses after recombination, whereas the astronomical objects of interest have masses on the order of $10^{12}$ solar masses or greater~\citep{cit:Huterer-book}, we can assume a length scale $\lambda\gg \lambda_\mathrm{J}$, which is equivalent to $c_\mathrm{s}\approx 0$. Thus, at late times, when $c_\mathrm{s}=0$ is a reasonably good approximation, Equation~\eqref{eq:mastereq-K} reduces to a yet simpler form~\citep{cit:Huterer-book}
\begin{equation}
\deriv[2]{\delta}{t}+2\,\frac{\dot{a}}{a}\deriv{\delta}{t}=4\pi G\,\overline{\rho}\,\delta,
\label{eq:mastereq-Huterer}
\end{equation}
which is the equation we will solve to obtain $\delta$. We have dropped the $\vect{k}$ subscript because Equation~\eqref{eq:mastereq-Huterer} does not distinguish between perturbations of different scales (different $|\vect{k}|$). For convenience, we also make the substitution ${\delta(t)=\delta_0\,D(t)}$, where $D(t)$ denotes the \textit{growth factor}, which is normalised to 1 at the present time. Substituting this expression into Equation~\eqref{eq:mastereq-Huterer}, rewriting the right-hand side using $\bar{\rho}=(3H_0^2/8\pi G)\Omega_\mathrm{m,0}a^{-3}$, and $a(t) = H_0t$, Equation~\eqref{eq:mastereq-Huterer} becomes 
\begin{equation}
    \deriv[2]{D}{a} + \frac{2}{a}\,\deriv{D}{a}=\frac{3}{2}\Omega_\mathrm{m,0}\,a^{-3}D(a).
    \label{eq:masterequation}
\end{equation}
Then, by performing a change of variables, ${y\equiv \sqrt{6\Omega_\mathrm{m,0}/a}=\sqrt{6\Omega_\mathrm{m,0}(1+z)}}$, Equation~\eqref{eq:masterequation} takes the form of a Bessel differential equation, yielding the general solution
\begin{equation}
D(y(z))=y(z)\left(C_1\,I_1(y(z))+C_2\,K_1(y(z))\right),
\label{eq:analyticalsol}
\end{equation}
where $I_i$ and $K_i$ are the $i$th order modified Bessel functions of the first and second kind, respectively. On the relevant interval, $a\in(0,1)$, $I_1\left(y\right)$ represents the decaying mode and so is neglected. Furthermore, in view of the normalisation condition placed on $D(a)$, one may write the growth factor as a function of redshift as 
\begin{equation}
D(z)=\frac{y(z)}{\sqrt{6\Omega_\mathrm{m,0}}K_1\left(\sqrt{6\Omega_{\mathrm{m,0}}}\right)}K_1(y(z)),
\label{eq:Dcoast}
\end{equation}
where the dependence on scale factor is replaced by redshift, as 
$f\sigma_8(z)$ is typically reported in terms of redshift. Having derived $D(z)$, an expression for $f\sigma_8(z)$ can be obtained from the formula~\citep{cit:Huterer}
\begin{equation}
f\sigma_8(a)=\sigma_{8,0}D(a)\deriv{\ln{D}}{\ln{a}},
\label{eq:fs8a}
\end{equation}
where ${\sigma_{8,0}\equiv \sigma_8(a=1)=\sigma_8(z=0)}$.
Recasting Equation~\eqref{eq:fs8a} in terms of redshift yields the final form of $f\sigma_8(z)$ in a coasting universe
\begin{equation}\begin{split}
f\sigma_8(z)&=-\frac{\sigma_{8,0}}{\sqrt{6\Omega_\mathrm{m,0}}K_1\left(\sqrt{6\Omega_{\mathrm{m,0}}}\right)}\cdot
\Biggl[ \frac{ y(z)K_1(y(z))}{2}\\&-\frac{y^2}{4}\biggl(K_0(y(z))+K_2(y(z))\biggr)\Biggr].
\end{split}
\label{eq:fitted}
\end{equation}

It ought to be noted that the current work is not the first instance of Equation~\eqref{eq:mastereq-Huterer} being solved without including components other than matter in $\overline{\rho}$. A solution provided by \citet{cit:Lemons-Peter} was analogous to Equation~\eqref{eq:analyticalsol}, as it was also a linear combination of the same modified Bessel functions. However, their approach, which treated Equation~\eqref{eq:masterequation} as an initial value problem, retained the decaying mode, whereas we consider it negligible. 

\section{Data and Analysis}
\label{sec:Data_and_Analysis}
The data utilised in this study were originally assembled by \citet{cit:Perivolaropoulos-Skara} to investigate, among other things, the discrepancy between values of $\Omega_\mathrm{m,0}$ and $\sigma_{8,0}$ derived from redshift space distortion and \textit{Planck}~\citep{cit:planck2018} data. The data compilation of \citet{cit:Perivolaropoulos-Skara} contains a total of sixty-six $f\sigma_8(z)$ values at various redshifts. However, many of these data points come from galaxy surveys whose galaxy samples overlap, which introduces correlations between data points and their uncertainties. A full covariance matrix is therefore not available for the whole dataset. As such, for the purposes of the present study, we only used a subset of thirty-five data points, which \citet{cit:Perivolaropoulos-Skara} identified as being largely uncorrelated, so that the corresponding covariance matrix could be approximated by a diagonal matrix.

As outlined by \citet{cit:Kazantzidis-Perivolaropoulos}, obtaining $f\sigma_8(z)$ values from redshift space distortion data necessitates assuming a fiducial cosmology to convert redshifts into distances, which makes these values model-dependent. Consequently, such data cannot be used to test models which differ from the one originally assumed. However, one can calculate the value $f\sigma_8(z)$ would have taken had one assumed a different fiducial model by multiplying the existing data points with a correction factor $q(z, \Omega_\mathrm{m,0}, \Omega^\mathrm{fid}_\mathrm{m,0})$~\citep{cit:Kazantzidis-Perivolaropoulos}, where $\Omega^\mathrm{fid}_\mathrm{m,0}$ corresponds to the present-day value of the matter density parameter in the fiducial cosmology, and $\Omega_\mathrm{m,0}$ is the corresponding value in the model under scrutiny. This is an approximation based on the Alcock-Paczy\'{n}ski effect~\citep{cit:APeffect}. As \citet{cit:Kazantzidis-Perivolaropoulos} note, one may find different forms for $q(z, \Omega_\mathrm{m,0}, \Omega^\mathrm{fid}_\mathrm{m,0})$ in the literature, but we will use the one employed in other recent studies (e.g. by \citet{cit:Velasquez-Toribio, cit:Li-correction,cit:Arjona, cit:Borges,cit:Deliduman}) 
\begin{equation}
    q(z, \Omega_\mathrm{m,0}, \Omega^\mathrm{fid}_\mathrm{m,0}) = \frac{H(z)D_\mathrm{A}(z)}{H^\mathrm{fid}(z)D^\mathrm{fid}_{\mathrm{A}}(z)},
    \label{eq:correction}
\end{equation}
where $D_{\mathrm{A}}(z)$ is the angular diameter distance. For each data point in our dataset, the fiducial model was a flat $\Lambda$CDM cosmology (for which the density parameter of radiation was considered negligible), with the specific values of the various parameters differing depending on the source publication. Thus, one may write $H^\mathrm{fid}(z)$ and $D^\mathrm{fid}_A(z)$ as
\begin{equation}
H^\mathrm{fid}(z)=H^\mathrm{fid}_0 \sqrt{\Omega^\mathrm{fid}_{\mathrm{m,0}}(1+z)^3+(1-\Omega^\mathrm{fid}_{\mathrm{m,0}})}
\label{eq:H-lcdm}
\end{equation}
and
\begin{equation}
D^\mathrm{fid}_\mathrm{A}(z)=\frac{c}{H^\mathrm{fid}_0}\frac{1}{1+z}\int_{0}^{z}\frac{H^\mathrm{fid}_0}{ H^\mathrm{fid}(z’)}\diff z’
\label{eq:Da-lcdm}
\end{equation}
respectively, while the relevant quantities for a coasting cosmology are
\begin{equation}
H(z)=H_0(1+z)
\label{eq:H-c}
\end{equation}
and 
\begin{equation}
D_\mathrm{A}(z)=\frac{c}{H_0}
\begin{cases*}
    \frac{\sinh{(\ln{(1+z)})}}{(1+z)},& $k=-1$\\
    \frac{\ln{(1+z)}}{(1+z)}, & $k=0$\\
    \frac{\sin{(\ln{(1+z)})}}{(1+z)}, & $k=+1$\\
\end{cases*}.
\label{eq:Da-c}
\end{equation}
Substituting Equations \eqref{eq:H-lcdm}-\eqref{eq:Da-c} into Equation~\eqref{eq:correction} makes it clear that $q$ does not depend on $H^\mathrm{fid}_0$ and $H_0$, as both parameters cancel.

The aim of this work is to determine whether coasting cosmologies are consistent with the growth data from redshift space distortion measurements in the redshift range $z\in (0, 2)$ and whether they provide a better fit compared to the $\Lambda$CDM model. Therefore, the dataset was homogenised with respect to a flat $\Lambda$CDM model with the baseline \textit{Planck} 2018 best-fit value~\citep{cit:planck2018} of $\Omega_{\mathrm{m,0}}$ $(0.3153)$, as well as three coasting models, by multiplying both the fiducial $f\sigma_8$ values and their uncertainties by the corresponding $q$. Note that by Equations \eqref{eq:H-c} and \eqref{eq:Da-c}, no prior knowledge of $\Omega_\mathrm{m,0}$ was required to perform this step for the coasting cases, and the fiducial $\Omega^{\mathrm{fid}}_\mathrm{m,0}$ values were listed by \citet{cit:Perivolaropoulos-Skara}, facilitating the computation of $q(z, \Omega_\mathrm{m,0}, \Omega^\mathrm{fid}_\mathrm{m,0})$.

\begin{figure*}
 \begin{subfigure}[]{1\columnwidth}
    \caption{Coasting Model, $k=-1$}
     \includegraphics[width=1.05\textwidth]{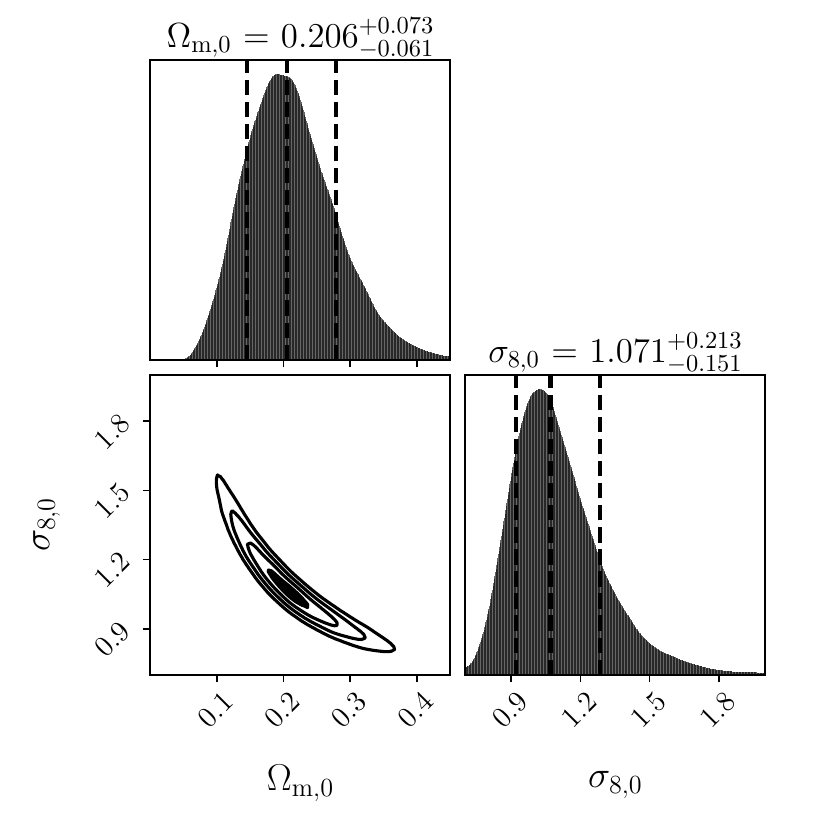}
     \label{fig:1a}
 \end{subfigure}
 \hfill
 \begin{subfigure}[]{1\columnwidth}
    \caption{Coasting Model, $k=0$}
     \includegraphics[width=1.05\textwidth]{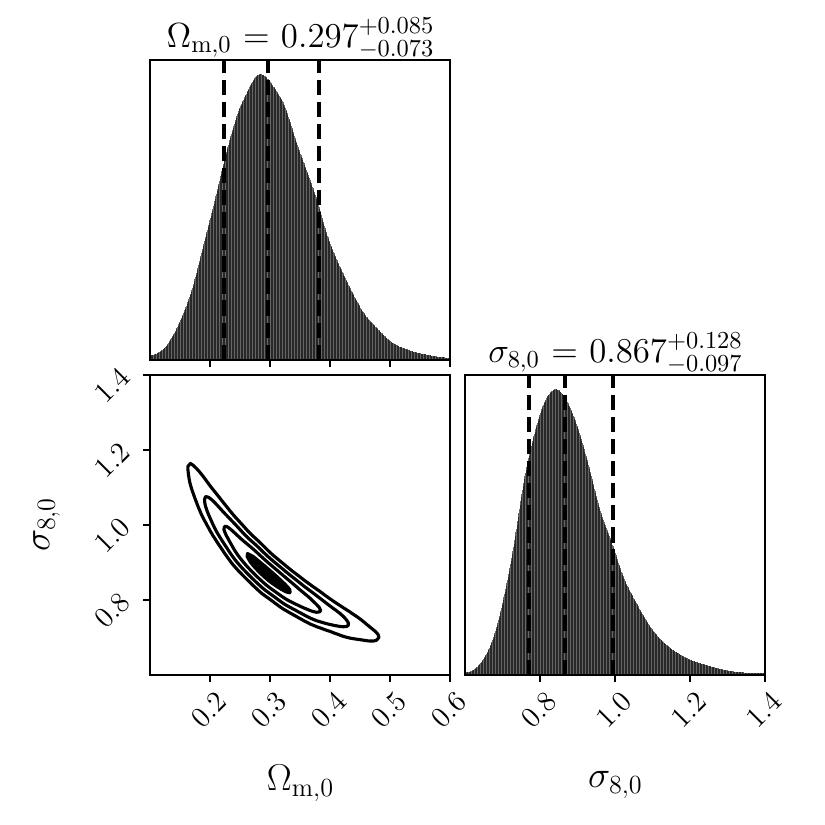}
     \label{fig:1b}
 \end{subfigure}
 
 \begin{subfigure}[]{1\columnwidth}
    \caption{Coasting Model, $k=+1$}
     \includegraphics[width=1.05\textwidth]{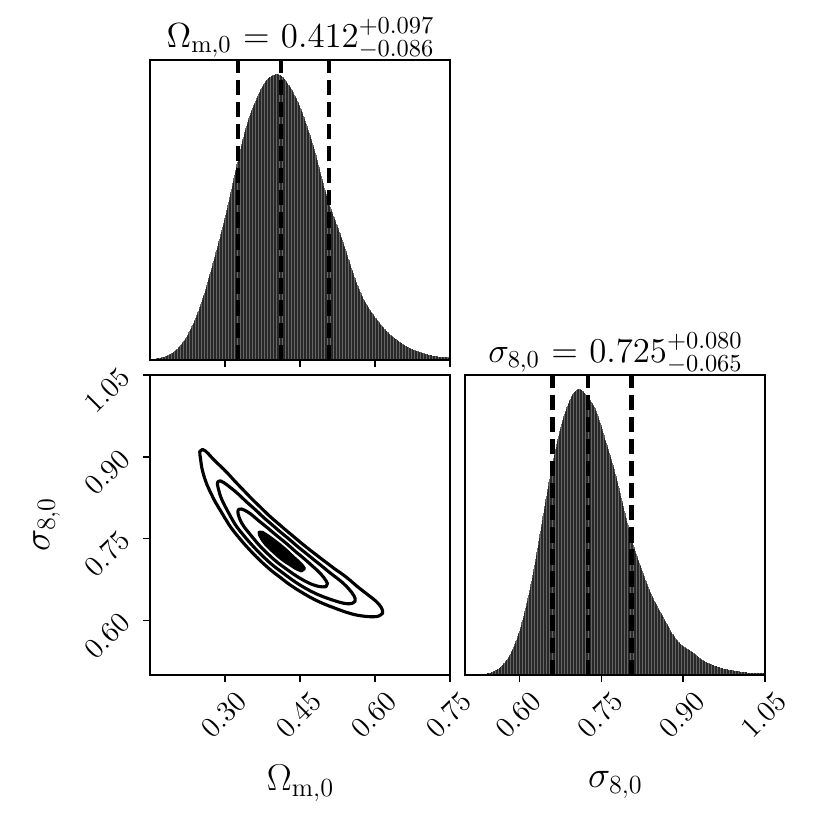}
     \label{fig:1c}
 \end{subfigure}
 \hfill
 \begin{subfigure}[]{1\columnwidth}
    \caption{$\Lambda$CDM Model}
     \includegraphics[width=1.05\textwidth]{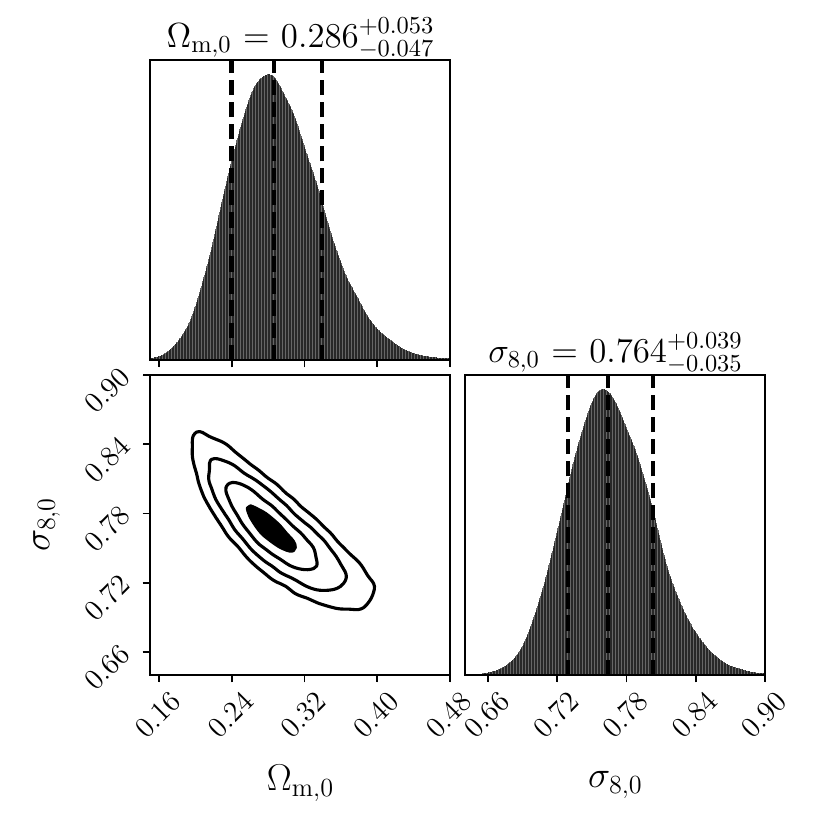}
     \label{fig:1d}
 \end{subfigure}

 \caption{Posterior distributions of $\Omega_\mathrm{m,0}$ and $\sigma_{8,0}$ for coasting models with different geometries (upper panels, lower left panel) and a flat $\Lambda$CDM cosmology (lower right panel), obtained from fitting thirty-five $f\sigma_8$ data points from \citet{cit:Perivolaropoulos-Skara}. The posteriors lie within the $(0, 2)$ interval of the prior distribution for both parameters. The programs used to obtain the posterior distributions are available in our public code repository~\citep{cit:Zenodo}.}
 \label{fig:posteriors}
\end{figure*}

Subsequent to recalibration, we fitted Equation~\eqref{eq:fitted} (for the coasting cases) to the data using the \texttt{dynesty} implementation~\citep{cit:Speagle,cit:Koposov} of nested sampling~\citep{cit:Skilling-2004,cit:Skilling-2006} with 10,000 live points, uniform sampling, multiple bounding ellipsoids~\citep{cit:Feroz-Hobson-Bridges}, and a convergence criterion of ${\Delta \log{\mathcal{Z}}\leq 0.001}$. The priors of $\Omega_\mathrm{m,0}$ and $\sigma_{8,0}$ were taken to be uniform distributions on the $(0, 2)$ interval. The lower bound is motivated by the fact that neither $\Omega_\mathrm{m,0}$ nor $\sigma_{8,0}$ can take negative values, as that would be unphysical. The upper bound was chosen to be 2 as neither parameter is expected to exceed it based on independent empirical determinations. The results in Figure~\ref{fig:posteriors} demonstrate that the posterior distributions lie within the $(0, 2)$ interval. This same approach was taken in fitting the corresponding $f\sigma_8(z)$ for the $\Lambda$CDM case, except that Equations \eqref{eq:H-lcdm} and \eqref{eq:Da-lcdm} were used for both the fiducial and the "target" cosmologies, with only the specific parameter values differing between the two. The theoretical $f\sigma_8(z)$ curve for the $\Lambda$CDM case was constructed from the following two functions. Firstly, the analytical expression for $\sigma_8(z)=\sigma_{8,0}D(z)$ was obtained by using the fact that ${D(a)=ag(a)/g(1)}$, where $g(a)$ is the growth suppression factor~\citep{cit:Huterer-book}. For a flat $\Lambda$CDM universe, at late times, $g(z)$ can be expressed as~\citep{cit:Mo-et-al-GFE}
\begin{equation}\begin{split}
    g(z)&= \frac{5}{2}\Omega_\mathrm{m}(z)\biggl[\Omega^{\frac{4}{7}}_\mathrm{m}(z)-\Omega_{\Lambda}(z)+\\
    &+\left(1+\frac{\Omega_\mathrm{m}(z)}{2}\right)\left(1+\frac{\Omega_{\Lambda}(z)}{70}\right)\biggr]^{-1}.
\end{split}\end{equation}
Secondly, there exists a convenient form of $f(z)$ for the $\Lambda$CDM universe, called the growth index parametrisation ${f(z)=\Omega^{\gamma}_\mathrm{m}(z)}$~\citep{cit:Wang-Steinhardt}, where ${\gamma=6/11}$ is implied by ${w_\mathrm{\Lambda}=-1}$~\citep{cit:Nesseris-Perivolaropoulos}.

Evaluation of the models’ consistency with the data and determination of model preference could not be performed using the reduced chi-squared statistic due to the nonlinearity of the models' dependencies on $\Omega_\mathrm{m,0}$ and $\sigma_{8,0}$, which prevents reliable determination of the number of degrees of freedom (for a didactic discussion of the reasons underlying this non-applicability, see~\citealt{cit:Andrae-Schulze-Hartung-Melchior}). Instead, we assessed model consistency with the data by computing the residuals of the fits, normalising them with recalibrated uncertainties, and subjecting them to the Anderson--Darling (AD) test for normality~\citep{cit:AD}. The AD test provides a quantitative measure of the extent to which the empirical cumulative distribution of the residuals differs from that expected for a standard normal distribution. A model with the correct parameter values would produce residuals which follow a standard normal distribution by definition~\citep{cit:Andrae-Schulze-Hartung-Melchior}. To complement the AD test, we performed posterior predictive checks as outlined by~\citet{cit:Gelman-Meng-Stern}.

When establishing model preference, we first drew upon the Bayesian evidence, $\mathcal{Z}$, computed by \texttt{dynesty} in the course of curve fitting, to quantify the preference through Bayes factors~\citep{cit:Bayes}. Following the recommendation of~\citet{cit:Andrae-Schulze-Hartung-Melchior}, we used leave-one-out cross-validation (LOO-CV) to calculate, using the framework set out by~\citet{cit:Vehtari-Gelman-Gabry} and detailed in Section~\ref{sec:Results}, a Bayesian LOO estimate of the expected log predictive densities (denoted as $\mathrm{elpd}_\mathrm{loo}$) of the models to assess their predictive performances. The sampling uncertainty in $\mathrm{elpd}_\mathrm{loo}$ was estimated by bootstrapping~\citep{cit:Andrae-Schulze-Hartung-Melchior}. The combination of the aforementioned metrics was used to assess both model viability and preference.

\section{Results and Discussion}
\label{sec:Results}

\begin{figure*}
 \begin{subfigure}[!h]{1\columnwidth}
     \includegraphics[width=1\textwidth]{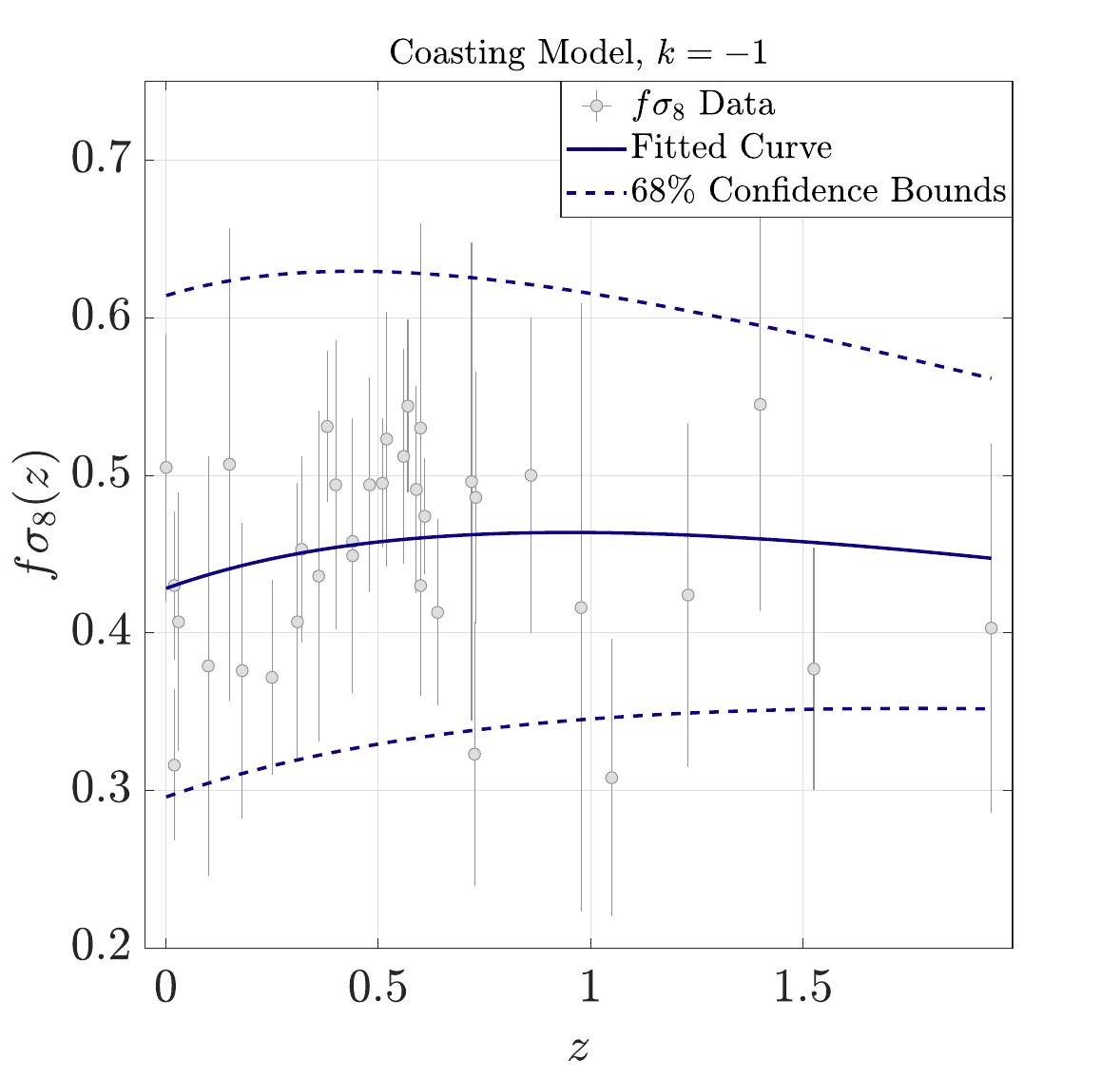}
     \label{fig:2a}
 \end{subfigure}
 \hfill
 \begin{subfigure}[!h]{1\columnwidth}
     \includegraphics[width=1\textwidth]{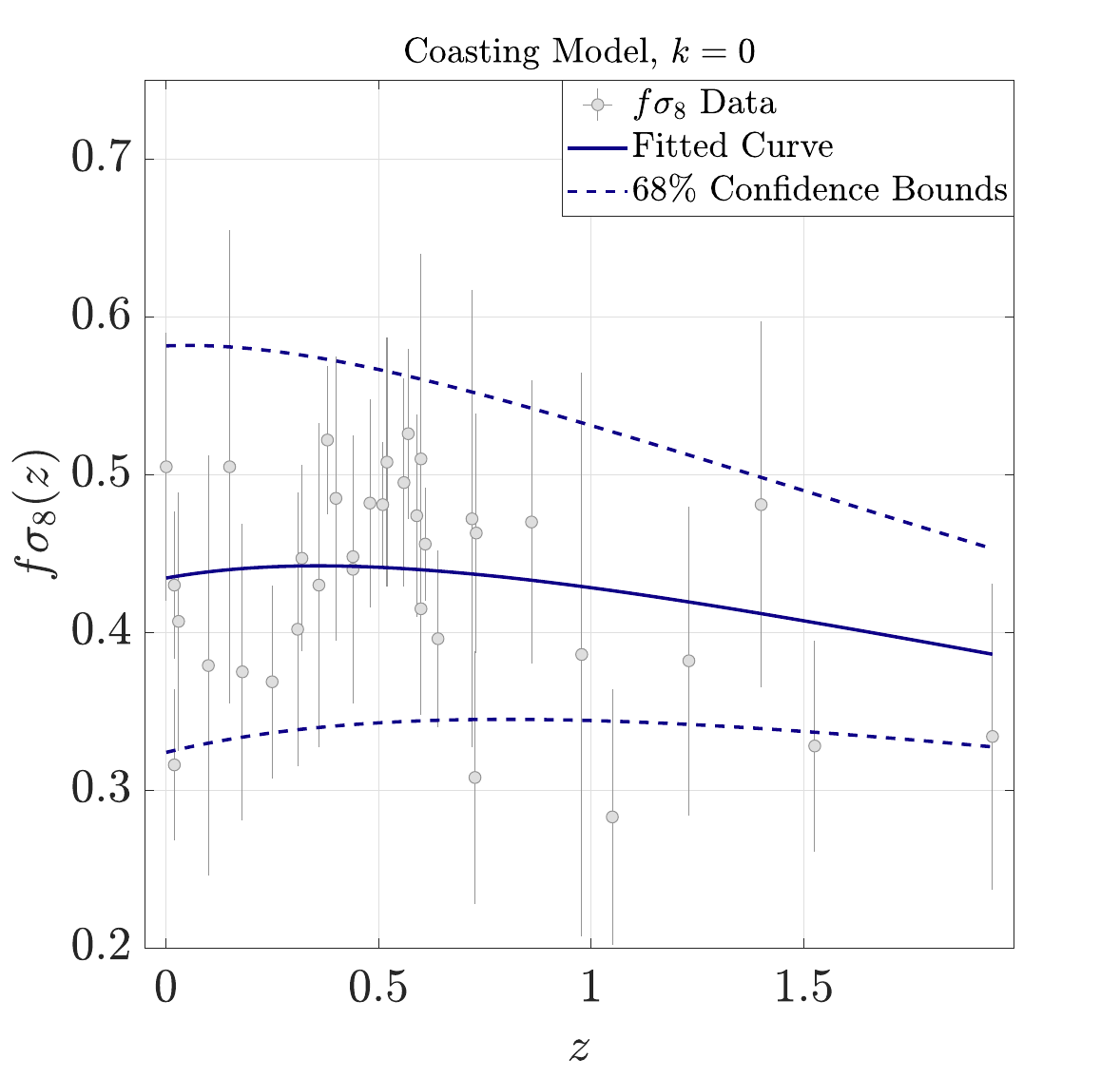}
     \label{fig:2b}
 \end{subfigure}
 
 \begin{subfigure}[!h]{1\columnwidth}
     \includegraphics[width=1\textwidth]{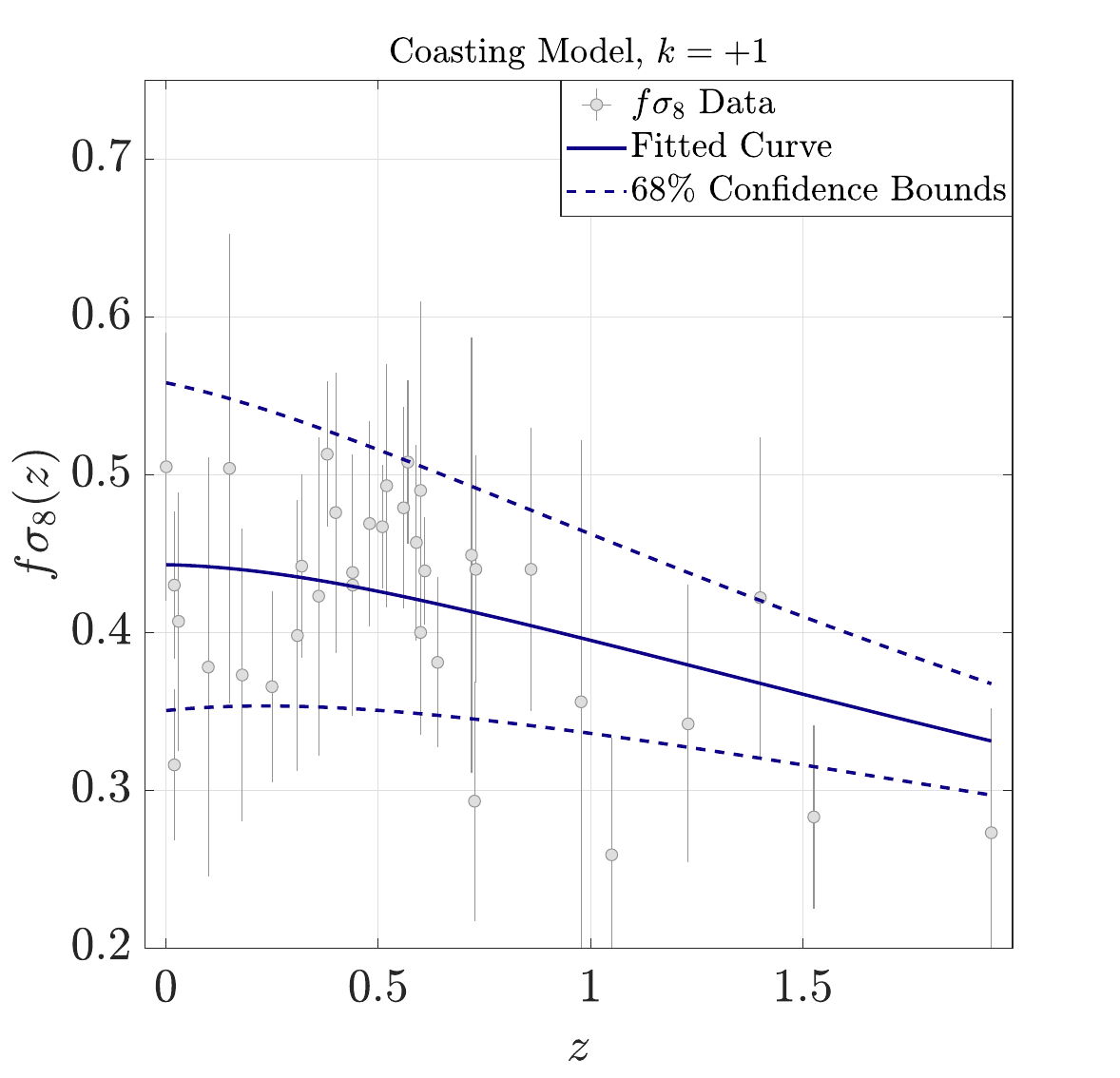}
     \label{fig:2c}
 \end{subfigure}
 \hfill
 \begin{subfigure}[!h]{1\columnwidth}
     \includegraphics[width=1\textwidth]{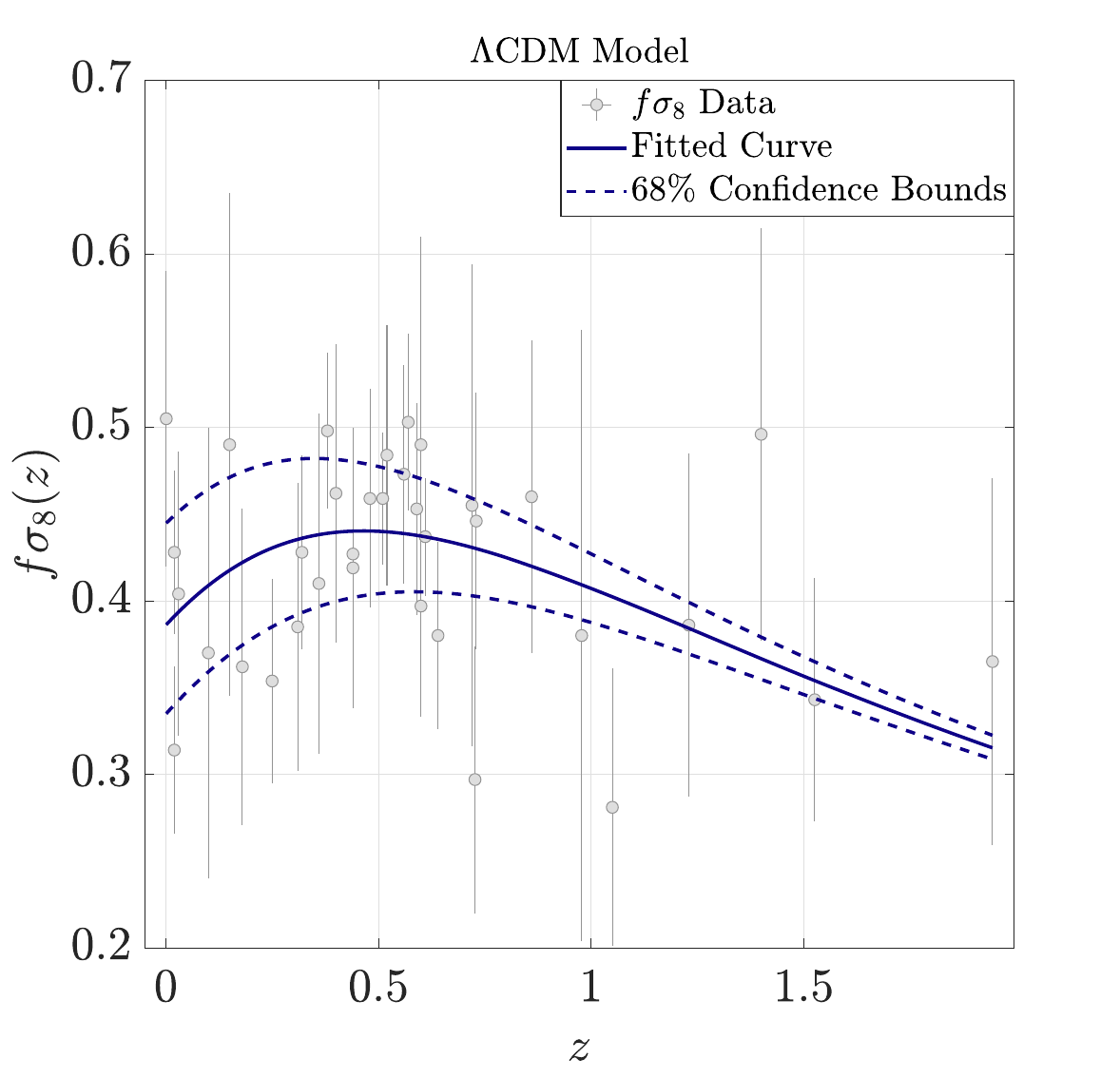}
     \label{fig:2d}
 \end{subfigure}

 \caption{Comparison of the density-weighted growth rate, $f\sigma_8$, for coasting models with different geometries (upper panels, lower left panel) and a $\Lambda$CDM cosmology (lower right panel). The fits are shown for redshift space distortion data collected by \citet{cit:Perivolaropoulos-Skara}, recalibrated for each model using a correction factor based on the Alcock-Paczy\'{n}ski effect~\citep{cit:APeffect}, as detailed by \citet{cit:Kazantzidis-Perivolaropoulos} and \citet{cit:Perivolaropoulos-Skara}. The dark blue lines represent the best-fit curves. Dashed lines indicate the 68\% confidence intervals. The dataset consists of thirty-five largely uncorrelated data points. The programs used to obtain the posterior distributions are available in our public code repository~\citep{cit:Zenodo}.}
 \label{fig:fitted}
\end{figure*}

We fitted Equation~\eqref{eq:fitted} to three coasting models, with curvature parameters ${k=\{-1, 0, +1\}}$. The best-fit values of parameters correspond to the median values, while the uncertainties are derived by taking the difference between the 84th and 16th percentiles and the median. Using these parameter values, we calculated the uncertainty-normalised residuals, denoted as $R$, for the recalibrated dataset. The uncertainty-normalised residuals are defined as 
\begin{equation}
    R = \frac{(f\sigma_8(z))_\mathrm{data} - (f\sigma_8(z))_\mathrm{model}}{ \sigma_\mathrm{data}},
\end{equation} where $(f\sigma_8(z))_\mathrm{data}$ is the observed value, $(f\sigma_8(z))_\mathrm{model}$ is the model-predicted value, and $\sigma_\mathrm{data}$ is the uncertainty in the data.
We then conducted an AD test to assess how well these residuals adhere to a standard normal distribution. Using the MATLAB implementation of the AD test~\citep{cit:matlab} with a 5\% significance level, the test failed to reject null hypothesis that $R$ follows a standard normal distribution for all three values of $k$. The test yielded \textit{p}-values of ${p=\{0.655,\, 0.713,\,0.778\}}$. The results of the fits are summarised in Table~\ref{tab:results} and illustrated in Figure~\ref{fig:fitted}.

\begin{figure}
    \centering
    \includegraphics[width=0.5\textwidth]{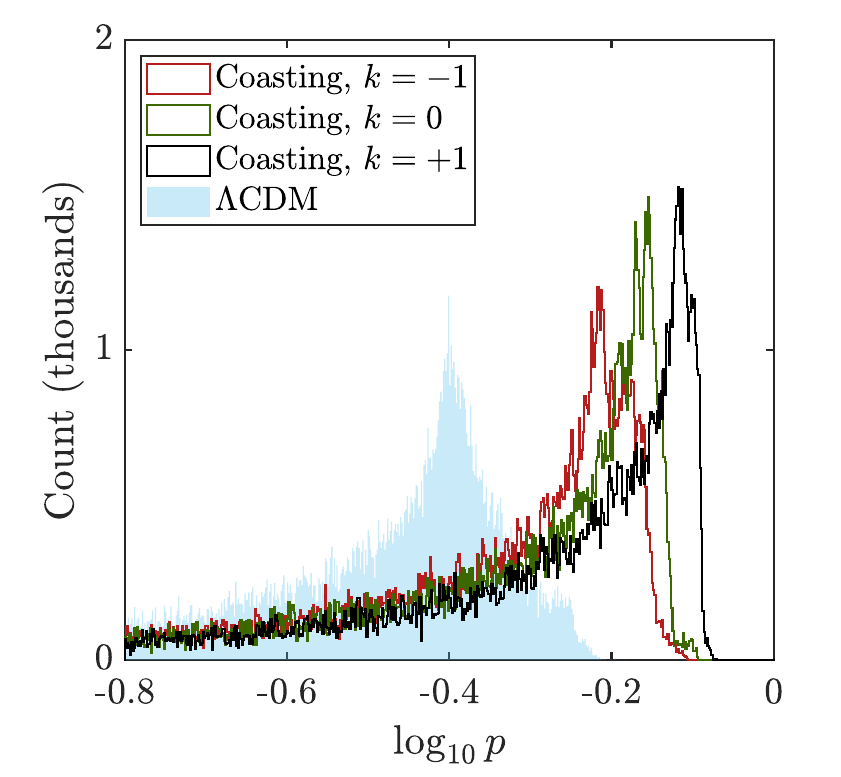}
    \caption{Histograms of AD-test $\log_{10}p$ values for one hundred and thirty-seven thousand realisations of the coasting and $\Lambda$CDM models, sampled from the joint posterior distribution of $\Omega_{\mathrm{m},0}$ and $\sigma_{8,0}$ obtained from fitting to thirty-five $f\sigma_8$ data points from \citet{cit:Perivolaropoulos-Skara}. More than 95\% of $p$-values exceeded 0.05 for all four models.}
    \label{fig:fig3}
\end{figure}

Repeating the analysis for the dataset homogenised with respect to the baseline \textit{Planck} 2018 $\Lambda$CDM cosmology, we obtained the following best-fit values: ${\Omega_\mathrm{m,0}=0.286^{+0.053}_{-0.046}}$ and ${\sigma_{8,0}=0.764^{+0.039}_{-0.035}}$, which deviate from the \textit{Planck} 2018 values by $0.55\sigma$ and $1.2\sigma$, respectively. The AD test failed to reject the null hypothesis of a standard normal distribution with a $p$-value of 0.416. These results suggest that both the concordance $\Lambda$CDM model and the coasting models are consistent with the available growth data from redshift space distortion measurements. It should also be noted that the coasting models generally yielded higher $p$-values than the flat $\Lambda$CDM model, as seen in Figure~\ref{fig:fig3}. The median $p$-value for the $\Lambda$CDM model is exceeded by ${\{71\%, 73\%, 77\%\}}$ of $p$-values for the ${k=\{-1, 0, +1\}}$ coasting models, respectively, while the maximum $p$-value is exceeded by ${\{12\%, 26\%, 39\%\}}$.

Mirroring the approach of \citet{cit:Perivolaropoulos-Skara}, we also fitted the models to the full dataset to assess whether consistency with the data is preserved and whether the best-fit parameters remain in agreement with those obtained from the subset. All four models remained consistent with the full dataset, though the $p$-values from the AD test were notably lower. The $p$-value for the coasting model with $k=+1$ was once again the largest, followed by those of the $k=0$ and $k=-1$ coasting models, and lastly that of the $\Lambda$CDM model. In terms of best-fit parameters, both $\Omega_\mathrm{m,0}$ and $\sigma_{8,0}$ differed by less than $0.5\sigma$ from their corresponding values obtained from the subset, across all models.

\begin{table*}

\caption{
Summary of the growth-data constraints and model-comparison diagnostics for the three coasting models and flat $\Lambda$CDM. The columns give the posterior constraints on $\Omega_{\mathrm{m,0}}$, $\sigma_{8,0}$, and $S_8$, the Anderson--Darling residual normality-test $p$-value, the log Bayes factor relative to each coasting model, the LOO-CV significance probability $p_k$, and the posterior predictive $p$-value $p_b$.
}
\label{tab:results}
\centering
\begin{tabular}{c|ccccccc}
\toprule
 Model & $\Omega_{\mathrm{m,0}}$  & $\sigma_{8,0}$                               & $S_8$                 & $p$    & $\log_{10}\mathcal{B}$  & $p_k$ & $p_b$         \\ \midrule
 Coasting, $k=-1$   & $0.206_{-0.061}^{+0.073}$ & $1.071_{-0.151}^{+0.213}$ & $0.890_{-0.024}^{+0.024}$ & 0.655 & 1.79 & 0.996 & $0.781$ \\
 Coasting, $k=0$   &  $0.297_{-0.073}^{+0.085}$ & $0.867_{-0.097}^{+0.128}$ & $0.865_{-0.024}^{+0.024}$  & 0.713 & 1.55 & 0.976 & $0.717$\\
 Coasting, $k=+1$& $0.412_{-0.086}^{+0.097}$ & $0.725_{-0.065}^{+0.080}$ & $0.850_{-0.026}^{+0.026}$ & 0.778 &  1.42  & 0.939   & $0.616$     \\ 
Flat $\Lambda$CDM& $0.286_{-0.047}^{+0.053}$ & $0.764_{-0.035}^{+0.039}$& $0.746_{-0.039}^{+0.041}$  & 0.416 &   0     & -      &  $0.926$      \\ \bottomrule
\end{tabular}
\end{table*}

In evaluating which model provides a better fit to the growth data, we used the base-10 logarithm of the ratio of Bayesian evidences, $\mathcal{Z}$, calculated by \texttt{dynesty}, referred to as the log Bayes factor, defined as ${\log_{10}{\mathcal{B}} \equiv \log_{10}(\mathcal{Z}_{\Lambda\mathrm{CDM}}/\mathcal{Z}_\mathrm{coasting})}$, as a first determination of model preference. For ${k=\{-1, 0, +1\}}$, we found ${\log_{10}{\mathcal{B}} = \{1.79,\, 1.55,\,1.42\}}$, respectively. Based on the scale devised by~\citet{cit:Bayes}, this indicates a strong preference for the $\Lambda$CDM model. We also examined the run-to-run variability of $\log_{10}{\mathcal{B}}$, by repeating the fit for each of the four models one hundred times. The mean values were $\log_{10}{\mathcal{B}}=\{1.79,\, 1.54,\,1.42\}$, and their standard deviations were uniformly $\sigma_{\log_{10}{\mathcal{B}}}=0.01$.

Furthermore, we also tested the prior sensitivity of $\log_{10}{\mathcal{B}}$. First, we decreased the prior volume by restricting $\Omega_\mathrm{m,0}$ to $(0, 1)$. We did not do the same for $\sigma_{8,0}$ because the $k=-1$ coasting model's $\sigma_{8,0}$ posterior extends to 2. The resulting log Bayes factors were $\log_{10}{\mathcal{B}} = \{1.78,\, 1.54,\,1.42\}$, which are nearly identical to the original values. Next, we tested whether using a $S_8\sim\mathcal{U}(0, 2)$ prior and $\sigma_{8,0}=S_8\sqrt{0.3/\Omega_\mathrm{m,0}}$ would impact the log Bayes factors.  We found ${\log_{10}{\mathcal{B}} = \{1.87,\, 1.55,\,1.34\}}
$. These results indicate that neither a restricted prior volume nor sampling in $S_8$ rather than $\sigma_{8,0}$ alters the level of preference for $\Lambda$CDM. We conducted a further robustness test by changing the likelihood functions to ones based on Student's $t$-distribution rather than a Gaussian distribution, with degrees of freedom parameters $\nu=\{5,10,20\}$ (corresponding to excess kurtoses of 6, 1, and 0.375, respectively). For these values of $\nu$, we obtained ${\log_{10}{\mathcal{B}} = \{1.51,\, 1.16,\,0.92\}}$, ${\log_{10}{\mathcal{B}} = \{1.60,\, 1.29,\,1.08\}}$, and ${\log_{10}{\mathcal{B}} = \{1.71,\, 1.40,\,1.23\}}$, respectively. Thus, it is apparent that as the departure from Gaussianity increases, log Bayes factors decrease. However, the $\Lambda$CDM model remains strongly preferred over the $k=\{-1, 0\}$ coasting models, whereas for $k=+1$ the preference reaches the border of substantial and strong ($\log_{10}{\mathcal{B}} = 1$) as $\nu$ decreases. 

Subsequently, we computed the $\mathrm{elpd}_\mathrm{loo}$ of each model to assess their predictive performances. The LOO predictive density of the data point $y_i$ ${(i\in [1,n])}$ is defined as~\citep{cit:Vehtari-Gelman-Gabry}
\begin{equation}
p(y_i|y_{-i})=\int p(y_i|\theta)p(\theta|y_{-i})\diff{\theta},
\label{eq:p_loo}
\end{equation}
where $\theta$ represents the model parameters, from which $\mathrm{elpd}_\mathrm{loo}$ is~\citep{cit:Vehtari-Gelman-Gabry}
\begin{equation}
\mathrm{elpd}_\mathrm{loo}=\sum_{i=1}^{n}\log{p(y_i|y_{-i})}.
\label{eq:elpd_loo}
\end{equation}
Since the dataset comprises thirty-five data points, ${n=35}$. Using the \texttt{resample\_equal} function in \texttt{dynesty}, we obtained a sample of the posterior distributions for both $\Omega_\mathrm{m,0}$ and $\sigma_{8,0}$. With a sample of size $S$ drawn for each left-out data point $y_i$, we can compute the contribution of $y_i$ to $\mathrm{elpd}_\mathrm{loo}$ ($\mathrm{elpd}_i$) using the formula~\citep{cit:Vehtari-Gelman-Gabry}
\begin{equation}
    \mathrm{elpd}_i=\log{\left(\frac{1}{S}\sum_{s=1}^S p(y_i|\theta^s)\right)}.
    \label{eq:elpd_i}
\end{equation}
The overall predictive performance measure is then given by $\mathrm{elpd}_\mathrm{loo}
=\sum_{i=1}^n\mathrm{elpd}_i$. For the ${k=\{-1, 0, +1\}}$ coasting models, we found ${\mathrm{elpd}_\mathrm{loo}=\{72.5,73.0,73.1\}}$, while for the $\Lambda$CDM model we found $\mathrm{elpd}_\mathrm{loo}=77.6$. As the value of $\mathrm{elpd}_\mathrm{loo}$ is largest for the $\Lambda$CDM model, we define $\Delta \mathrm{elpd}_k\equiv \mathrm{elpd}_{\mathrm{loo},\Lambda\mathrm{CDM}}-\mathrm{elpd}_{\mathrm{loo},k}$, where $k$ corresponds to the curvature parameter of a given coasting model. As an approximate first estimate of the significance of these differences, we computed the normal-standardised elpd differences
\begin{equation}
Z_k =
\frac{|\Delta\mathrm{elpd}_k|}
{\sqrt{n\,\mathrm{Var}_i(\delta_{i,k})}},
\label{eq:z_elpd}
\end{equation}
where ${\delta_{i,k}\equiv \mathrm{elpd}_{i,\Lambda\mathrm{CDM}}-\mathrm{elpd}_{i,k}}$, and $\mathrm{Var}_i(\delta_{i,k})$ is the empirical variance of these pointwise differences over the data index $i$. For the ${k=\{-1,0,+1\}}$ coasting models, we obtain ${Z_k=\{2.72,2.06,1.64\}}$. These values suggest that the preference for $\Lambda$CDM over the ${k=-1}$ and ${k=0}$ coasting models is approximately at or above the $2\sigma$ level, while the preference over the ${k=+1}$ model is weaker.

To obtain a more robust assessment of the statistical significance of $\Delta \mathrm{elpd}_k$, the uncertainties of these differences were estimated using non-parametric bootstrapping, as suggested by~\citet{cit:Vehtari-Gelman-Gabry}. We generated $10^6$ bootstrap replicates of ${\delta^{b}_{i,k}\equiv \mathrm{elpd}_{i,\Lambda\mathrm{CDM}}-\mathrm{elpd}_{i,k}}$, and for each replicate computed the sum $\tilde{\delta}^b_k=\sum_{i=1}^n \delta^{b}_{i,k}$. From the resulting empirical distribution of $\tilde{\delta}^b_k$, we obtained its 2.5th and 97.5th percentiles, which define the 95\% confidence interval, and the one-sided probability $p_{k}\equiv P(\Delta \mathrm{elpd}_k>0)$. A difference $\Delta \mathrm{elpd}_k$ was considered significant if the entire 95\% confidence interval lay above zero. This condition is equivalent to requiring ${1 - p_k < 0.025}$, or ${p_k > 0.975}$. For the ${k=\{-1, 0, +1\}}$ coasting models, ${p_k=\{0.996,0.976,0.939\}}$, indicating $\Delta \mathrm{elpd}_k=0$ fell outside of the 95\% confidence interval only for ${k=\{-1, 0\}}$. Thus, using the bootstrap confidence intervals as our primary significance criterion, we conclude that the excess predictive power of the $\Lambda$CDM model over the coasting models is statistically significant for ${k=\{-1, 0\}}$, but not for ${k = +1}$.

As a complementary diagnostic, we performed a posterior predictive check to assess the absolute fit and uncertainty calibration of each model, following the methodology of~\citet{cit:Gelman-Meng-Stern}. For a given model, the posterior predictive $p$-value is defined as~\citep{cit:Gelman-Meng-Stern}
\begin{equation}
\begin{split}
p_b(y)= \int &P\left(T\left(y^{\mathrm{rep}}, \theta\right)\ge T\left(y, \theta\right)|\theta\right)\\
&\cdot P(\theta |y)\diff{\theta},
\end{split}
\label{eq:pb_integral}
\end{equation}
where $y$ denotes the observed data, $y^{\mathrm{rep}}$ denotes replicated datasets drawn from the posterior predictive distribution, and $T(y,\theta)$ is a chosen discrepancy statistic. In this work, the replicated data are generated as
\begin{equation}
y^{\mathrm{rep}}_i\sim\mathcal{N}\left(f\sigma_8(z_i,\theta),\sigma_i^2\right),
\end{equation}
and we use a $\chi^2$-like discrepancy,
\begin{equation}
    T(y^{(\mathrm{rep})}, \theta) =
    \sum_{i=1}^{n}
    \frac{\left(y^{(\mathrm{rep})}_i - f\sigma_8(z_i, \theta)\right)^2}{\sigma_i^2}.
\end{equation}
Here, $\sigma_i^2$ denotes the variance used in the likelihood for each data point.

We evaluate Equation~\eqref{eq:pb_integral} by Monte Carlo integration. For each model, we draw $S$ samples from the resampled posterior, and for each posterior draw we generate $R$ replicated datasets. The posterior predictive $p$-value is then estimated as
\begin{equation} 
\begin{split} 
p_b= \frac{1}{S}&\sum_{s=1}^{S}\biggl[\frac{1}{R}\\&\sum_{r=1}^{R} I\left\{T\left(y^{\mathrm{rep}}_{r,s}, \theta_s\right)\ge T\left(y, \theta_s\right)\right\}\biggr], 
\end{split} 
\label{eq:pb_montecarlo} 
\end{equation}
where $I$ is the indicator function.

For this discrepancy, values of $p_b$ close to 0.5 indicate that the observed weighted residual scatter is typical of the replicated datasets. Values of $p_b\ll 0.5$ indicate that the observed data are more discrepant from the model than expected under the posterior predictive distribution, while values of $p_b\gg 0.5$ indicate that the observed data are less discrepant than typical replicated datasets. The latter case should not be interpreted as direct evidence of overfitting in the predictive sense; rather, it indicates that the residual scatter is small relative to the assumed predictive variance, and therefore may signal overestimated uncertainties, correlations not captured by the likelihood, or other forms of uncertainty miscalibration.

For the purposes of this work, we set ${R=10^3}$ and ${S=10^4}$. For the ${k=\{-1,0,+1\}}$ coasting models, we find ${p_b=\{0.781,0.717,0.616\}}$, while for the $\Lambda$CDM model we find ${p_b=0.926}$. The standard Monte Carlo errors of these estimates are approximately 0.001 for each model. Since all values exceed 0.5, the observed weighted residual scatter is smaller than the scatter typically produced by posterior predictive replications. This effect is strongest for $\Lambda$CDM. Thus, the posterior predictive check does not constitute a separate model-selection criterion, nor do we interpret it as a direct test of predictive overfitting. Rather, it qualifies the evidence and LOO-CV results by showing that the model with the strongest support, $\Lambda$CDM, also has the most overdispersed posterior predictive distribution relative to the observed residual scatter. In other words, $\Lambda$CDM provides the best fit and predictive performance among the models tested, but its predictive uncertainties are conservative with respect to this discrepancy statistic. The coasting models are less favoured by the evidence and LOO-CV comparisons, but their posterior predictive discrepancies are closer to those obtained from replicated datasets.

Subsequent to performing model comparisons, we turned our attention to the present value of the weighted amplitude of matter fluctuations, $S_8$. The aim of this analysis was not to revisit the observational status of the growth tension in $\Lambda$CDM, but rather to determine whether an analogous discrepancy appears when the redshift-space distortion data are interpreted within quasi-linear coasting cosmologies. This distinction is important because $S_8$ is not a directly observed quantity: its inferred value depends on the assumed expansion history, matter density, and growth model~\citep{cit:Sabogal-S8}. Thus, a comparison between the CMB-inferred value of $S_8$ and the value reconstructed from late-time LSS data provides a model-dependent consistency test. The question of interest is whether the late-time growth data, when homogenised and fitted under the coasting assumption, remain discrepant with the reference \textit{Planck} 2018 value, $S_8=0.832\pm0.013$, obtained under flat $\Lambda$CDM~\citep{cit:planck2018}.

Regarding the redshift-space distortion data, we obtained $S_8=0.746^{+0.041}_{-0.039}$ when homogenising the dataset with respect to the baseline \textit{Planck} 2018 value of $\Omega_{\mathrm{m},0}$. This corresponds to a discrepancy of $\Delta S_8^{\Lambda\mathrm{CDM}}=2.00\sigma$ relative to the \textit{Planck} 2018 value, while remaining consistent with the DES-Y3 and KiDS-1000 results. In contrast, fitting Equation~\eqref{eq:fitted} yielded ${S_8=\{0.890^{+0.024}_{-0.024},\,0.865^{+0.024}_{-0.024},\,0.850^{+0.026}_{-0.026}\}}$
for the $k=\{-1,0,+1\}$ quasi-linear coasting models, respectively. These correspond to discrepancies of
${\Delta S_8^{\mathrm{Coasting}}=\{2.12\sigma,\,1.21\sigma,\,0.62\sigma\}}$
with respect to the same \textit{Planck} 2018 reference value. Thus, relative to this particular CMB-inferred $\Lambda$CDM baseline, the flat and positively curved quasi-linear coasting models reduce the apparent discrepancy in $S_8$, while the negatively curved case does not.

This result should not be interpreted as a model-independent resolution of the growth tension. The \textit{Planck} value used here is itself inferred within flat $\Lambda$CDM, and a fully self-consistent test would require computing the CMB anisotropy spectrum, matter transfer function, and corresponding value of $S_8$ within each quasi-linear coasting model. Nevertheless, because the transition to coasting expansion is assumed to occur after recombination, comparison with the \textit{Planck} $\Lambda$CDM value provides a useful reference test of whether the late-time growth data become more or less compatible with the standard early-universe normalisation. This caveat is stronger for strictly linearly coasting cosmologies, which do not share the same pre-recombination expansion history.

\section{Conclusion}
\label{sec:Conclusions}

In this work, we investigated the consistency of quasi-linear coasting cosmologies with late-time measurements of the growth of large-scale structure. Starting from the standard fluid treatment of matter perturbations, and assuming only that the scale factor is linear in cosmic time over the redshift interval probed by the data, we derived a closed-form expression for the growth factor, $D(z)$, and the corresponding density-weighted growth rate, $f\sigma_8(z)$. This expression was then fitted to a subset of thirty-five largely uncorrelated redshift-space-distortion measurements, recalibrated for coasting cosmologies with curvature parameters ${k={-1,0,+1}}$ using an Alcock--Paczy'{n}ski correction. For comparison, the same analysis was also performed for a flat $\Lambda$CDM cosmology using the baseline \textit{Planck} 2018 value of $\Omega_{\mathrm{m},0}$.

We found that all three coasting models, as well as the flat $\Lambda$CDM model, are statistically consistent with the available $f\sigma_8$ data. In each case, the Anderson--Darling test failed to reject the hypothesis that the uncertainty-normalised residuals are drawn from a standard normal distribution. Thus, late-time growth measurements do not rule out quasi-linear coasting expansion over the redshift range $z\in(0,2)$. With regards to model preference, the Bayesian evidences strongly favour $\Lambda$CDM over all three coasting models, and this conclusion was found to be robust against changes in prior volume, parametrisation, and likelihood shape. Leave-one-out cross-validation similarly favours $\Lambda$CDM in terms of predictive performance, although the preference is statistically significant only for the $k=-1$ and $k=0$ coasting models, and not for the positively curved case. The posterior predictive checks further indicate that the observed weighted residual scatter is smaller than that typically generated by the posterior predictive distribution, most strongly for $\Lambda$CDM. This does not provide a separate model-selection criterion, but it qualifies the evidence and LOO-CV results by showing that the best-supported model also has the most conservative predictive uncertainty calibration with respect to the discrepancy statistic used here.

We also examined whether an analogue of the $S_8$ discrepancy appears when the same growth data are interpreted within quasi-linear coasting cosmologies. For the flat $\Lambda$CDM fit, the reconstructed value of $S_8$ differs from the baseline \textit{Planck} 2018 value by approximately $2.00\sigma$. For the coasting models with ${k={-1,0,+1}}$, the corresponding discrepancies are ${{2.12\sigma,1.21\sigma,0.62\sigma}}$, respectively. Thus, relative to this reference $\Lambda$CDM CMB normalisation, the flat and positively curved coasting models reduce the apparent $S_8$ discrepancy, whereas the negatively curved model does not. This result should not be interpreted as a model-independent resolution of the growth tension, since the reference \textit{Planck} value is itself inferred within flat $\Lambda$CDM. A fully self-consistent test would require deriving the CMB anisotropy spectrum, matter transfer function, and early-time normalisation of perturbations within each quasi-linear coasting model.

The present analysis therefore supports a limited but useful conclusion: quasi-linear coasting cosmologies are compatible with current low-redshift redshift-space-distortion growth data, but they are not generally preferred over $\Lambda$CDM by model-comparison diagnostics. Future work should extend this test by incorporating a full covariance treatment of overlapping growth measurements, jointly fitting expansion and growth probes, and modelling the transition from the early $\Lambda$CDM-like phase to the late-time coasting regime. Most importantly, a self-consistent calculation of CMB and matter-power-spectrum observables in quasi-linear coasting cosmologies is required before their implications for the growth tension can be assessed decisively.

\backmatter

\section*{Acknowledgements}
The authors thank Leandros Perivolaropoulos for his correspondence on the correction factor used to recalibrate the $f\sigma_8$ dataset, Adrienn Pataki for her assistance and feedback regarding the software used in this investigation, Ioannis Pantos and Leandros Perivolaropoulos for the discussions on the current status of the $S_8$ tension, and Moncy V. John and Jackson Levi Said for their bibliographical suggestions. 

\section*{Declarations}
\subsection*{Funding}
This project has received funding from the HUN-REN Hungarian Research Network and was also supported by the Nemzeti Kutatási, Fejlesztési és Innovációs Hivatal (NKFIH) excellence grant TKP2021-NKTA-64.

\subsection*{Competing Interests}
The authors have no competing interests to declare.

\subsection*{Data and Code Availability}
The data and codes used in this work are freely retrievable from the authors' Zenodo repository at \href{https://doi.org/10.5281/zenodo.15548205}{doi:10.5281/zenodo.15548205}.

\bibliography{sn-bibliography}

@ARTICLE{cit:Bull,
       author = {{Bull}, Philip and {Akrami}, Yashar and {Adamek}, Julian and {Baker}, Tessa and {Bellini}, Emilio and {Beltr{\'a}n Jim{\'e}nez}, Jose and {Bentivegna}, Eloisa and {Camera}, Stefano and {Clesse}, S{\'e}bastien and {Davis}, Jonathan H. and {Di Dio}, Enea and {Enander}, Jonas and {Heavens}, Alan and {Heisenberg}, Lavinia and {Hu}, Bin and {Llinares}, Claudio and {Maartens}, Roy and {M{\"o}rtsell}, Edvard and {Nadathur}, Seshadri and {Noller}, Johannes and {Pasechnik}, Roman and {Pawlowski}, Marcel S. and {Pereira}, Thiago S. and {Quartin}, Miguel and {Ricciardone}, Angelo and {Riemer-S{\o}rensen}, Signe and {Rinaldi}, Massimiliano and {Sakstein}, Jeremy and {Saltas}, Ippocratis D. and {Salzano}, Vincenzo and {Sawicki}, Ignacy and {Solomon}, Adam R. and {Spolyar}, Douglas and {Starkman}, Glenn D. and {Steer}, Dani{\`e}le and {Tereno}, Ismael and {Verde}, Licia and {Villaescusa-Navarro}, Francisco and {von Strauss}, Mikael and {Winther}, Hans A.},
        title = "{Beyond {\ensuremath{\Lambda}} CDM: Problems, solutions, and the road ahead}",
      journal = {Physics of the Dark Universe},
         year = 2016,
        month = jun,
       volume = {12},
        pages = {56-99},
          doi = {10.1016/j.dark.2016.02.001},
archivePrefix = {arXiv},
       eprint = {1512.05356},
 primaryClass = {astro-ph.CO},
       adsurl = {https://ui.adsabs.harvard.edu/abs/2016PDU....12...56B}
}

@ARTICLE{cit:Turner,
       author = {{Turner}, Michael S.},
        title = "{The Road to Precision Cosmology}",
      journal = {Annual Review of Nuclear and Particle Science},
         year = 2022,
        month = sep,
       volume = {72},
        pages = {1-35},
          doi = {10.1146/annurev-nucl-111119-041046},
archivePrefix = {arXiv},
       eprint = {2201.04741},
 primaryClass = {astro-ph.CO},
       adsurl = {https://ui.adsabs.harvard.edu/abs/2022ARNPS..72....1T}
}

@ARTICLE{cit:Casado,
       author = {{Casado}, Juan},
        title = "{Linear expansion models vs. standard cosmologies: a critical and historical overview}",
      journal = {Astrophysics and Space Science},
         year = 2020,
        month = jan,
       volume = {365},
       number = {1},
          eid = {16},
        pages = {16},
          doi = {10.1007/s10509-019-3720-z},
       adsurl = {https://ui.adsabs.harvard.edu/abs/2020Ap&SS.365...16C}
}

@ARTICLE{cit:Efstathiou,
       author = {{Efstathiou}, George},
        title = "{Challenges to the {\ensuremath{\Lambda}} CDM cosmology}",
      journal = {Philosophical Transactions of the Royal Society of London Series A},
         year = 2025,
       volume = {383},
       number = {2290},
          eid = {20240022},
        pages = {20240022},
          doi = {10.1098/rsta.2024.0022},
archivePrefix = {arXiv},
       eprint = {2406.12106},
 primaryClass = {astro-ph.CO},
       adsurl = {https://ui.adsabs.harvard.edu/abs/2025RSPTA.38340022E}
}

@ARTICLE{cit:Raffai,
       author = {{Raffai}, Peter and {P{\'a}lfi}, M{\'a}ria and {D{\'a}lya}, Gergely and {Gray}, Rachel},
        title = "{Constraints on Coasting Cosmological Models from Gravitational-wave Standard Sirens}",
      journal = {Astrophysical Journal},
         year = 2024,
        month = jan,
       volume = {961},
       number = {1},
          eid = {17},
        pages = {17},
          doi = {10.3847/1538-4357/ad1035},
archivePrefix = {arXiv},
       eprint = {2310.16556},
 primaryClass = {astro-ph.CO},
       adsurl = {https://ui.adsabs.harvard.edu/abs/2024ApJ...961...17R}
}

@ARTICLE{cit:Perivolaropoulos,
       author = {{Perivolaropoulos}, L. and {Skara}, F.},
        title = "{Challenges for {\ensuremath{\Lambda}}CDM: An update}",
      journal = {New Astronomy Review},
         year = 2022,
        month = dec,
       volume = {95},
          eid = {101659},
        pages = {101659},
          doi = {10.1016/j.newar.2022.101659},
archivePrefix = {arXiv},
       eprint = {2105.05208},
 primaryClass = {astro-ph.CO},
       adsurl = {https://ui.adsabs.harvard.edu/abs/2022NewAR..9501659P}
}

@ARTICLE{cit:Kolb,
       author = {{Kolb}, Edward W.},
        title = "{A Coasting Cosmology}",
      journal = {Astrophysical Journal},
         year = 1989,
        month = sep,
       volume = {344},
        pages = {543},
          doi = {10.1086/167825},
       adsurl = {https://ui.adsabs.harvard.edu/abs/1989ApJ...344..543K}
}

@ARTICLE{cit:Rhct-Structure,
       author = {{Melia}, Fulvio},
        title = "{The linear growth of structure in the R$_{h}$ = ct universe}",
      journal = {Monthly Notices of the Royal Astronomical Society},
         year = 2017,
        month = jan,
       volume = {464},
       number = {2},
        pages = {1966-1976},
          doi = {10.1093/mnras/stw2493},
archivePrefix = {arXiv},
       eprint = {1609.08576},
 primaryClass = {astro-ph.CO},
       adsurl = {https://ui.adsabs.harvard.edu/abs/2017MNRAS.464.1966M}
}

@ARTICLE{cit:Huterer,
       author = {{Huterer}, Dragan},
        title = "{Growth of cosmic structure}",
      journal = {Astronomy and Astrophysics Review},
         year = 2023,
        month = dec,
       volume = {31},
       number = {1},
          eid = {2},
        pages = {2},
          doi = {10.1007/s00159-023-00147-4},
archivePrefix = {arXiv},
       eprint = {2212.05003},
 primaryClass = {astro-ph.CO},
       adsurl = {https://ui.adsabs.harvard.edu/abs/2023A&ARv..31....2H}
}

@BOOK{cit:Mo-et-al-GFE,
       author = {{Mo}, Houjun and {van den Bosch}, Frank C. and {White}, Simon},
        title = "Galaxy Formation and Evolution",
         year = 2010,
        publisher = "Cambridge University Press",
doi={10.1017/CBO9780511807244}
}

@ARTICLE{cit:Perivolaropoulos-Skara,
       author = {{Skara}, F. and {Perivolaropoulos}, L.},
        title = "{Tension of the E$_{G}$ statistic and redshift space distortion data with the Planck-{\ensuremath{\Lambda}} CDM model and implications for weakening gravity}",
      journal = {Physical Review D},
         year = 2020,
        month = mar,
       volume = {101},
       number = {6},
          eid = {063521},
        pages = {063521},
          doi = {10.1103/PhysRevD.101.063521},
archivePrefix = {arXiv},
       eprint = {1911.10609},
 primaryClass = {astro-ph.CO},
       adsurl = {https://ui.adsabs.harvard.edu/abs/2020PhRvD.101f3521S}
}

@ARTICLE{cit:Lemons-Peter,
       author = {{Lemons}, D.~S. and {Peter}, W.},
        title = "{Gravitational instability in a coasting universe}",
      journal = {Astronomy and Astrophysics},
         year = 1992,
        month = nov,
       volume = {265},
       number = {2},
        pages = {373},
       adsurl = {https://ui.adsabs.harvard.edu/abs/1992A&A...265..373L}
}

@ARTICLE{cit:Planck-power_spectrum,
       author = {{Planck Collaboration} },
        title = "{Planck 2018 results. I. Overview and the cosmological legacy of Planck}",
      journal = {Astronomy and Astrophysics},
         year = 2020,
        month = sep,
       volume = {641},
          eid = {A1},
        pages = {A1},
          doi = {10.1051/0004-6361/201833880},
archivePrefix = {arXiv},
       eprint = {1807.06205},
 primaryClass = {astro-ph.CO},
       adsurl = {https://ui.adsabs.harvard.edu/abs/2020A&A...641A...1P}
}

@ARTICLE{cit:APeffect,
       author = {{Alcock}, C. and {Paczy\'{n}ski}, B.},
        title = "{An evolution free test for non-zero cosmological constant}",
      journal = {Nature},
         year = 1979,
        month = oct,
       volume = {281},
        pages = {358},
          doi = {10.1038/281358a0},
       adsurl = {https://ui.adsabs.harvard.edu/abs/1979Natur.281..358A}
}

@ARTICLE{cit:planck2018,
       author = {{Planck Collaboration}},
        title = "{Planck 2018 results. VI. Cosmological parameters}",
      journal = {Astronomy and Astrophysics},
         year = 2020,
        month = sep,
       volume = {641},
          eid = {A6},
        pages = {A6},
          doi = {10.1051/0004-6361/201833910},
archivePrefix = {arXiv},
       eprint = {1807.06209},
 primaryClass = {astro-ph.CO},
       adsurl = {https://ui.adsabs.harvard.edu/abs/2020A&A...641A...6P}
}

@ARTICLE{cit:Andrae-Schulze-Hartung-Melchior,
       author = {{Andrae}, Rene and {Schulze-Hartung}, Tim and {Melchior}, Peter},
        title = "{Dos and don'ts of reduced chi-squared}",
      journal = {arXiv e-prints},
         year = 2010,
        month = dec,
          eid = {arXiv:1012.3754},
        pages = {arXiv:1012.3754},
          doi = {10.48550/arXiv.1012.3754},
archivePrefix = {arXiv},
       eprint = {1012.3754},
 primaryClass = {astro-ph.IM},
       adsurl = {https://ui.adsabs.harvard.edu/abs/2010arXiv1012.3754A}
}

@ARTICLE{cit:Guth,
       author = {{Guth}, Alan H.},
        title = "{Inflationary universe: A possible solution to the horizon and flatness problems}",
      journal = {Physical Review D},
         year = 1981,
        month = jan,
       volume = {23},
       number = {2},
        pages = {347-356},
          doi = {10.1103/PhysRevD.23.347},
       adsurl = {https://ui.adsabs.harvard.edu/abs/1981PhRvD..23..347G}
}

@ARTICLE{cit:Sabogal-S8,
       author = {{Sabogal}, Miguel A. and {Silva}, Emanuelly and {Nunes}, Rafael C. and {Kumar}, Suresh and {Di Valentino}, Eleonora and {Giar{\`e}}, William},
        title = "{Quantifying $S_8$ tension and evidence for interacting dark energy from redshift-space distortion measurements}",
      volume={110},
   ISSN={2470-0029},
   DOI={10.1103/physrevd.110.123508},
   number={12},
   journal={Physical Review D},
   publisher={American Physical Society (APS)},
   year={2024},
   month=dec }

@ARTICLE{cit:Kazantzidis-Perivolaropoulos,
       author = {{Kazantzidis}, Lavrentios and {Perivolaropoulos}, Leandros},
        title = "{Evolution of the f {\ensuremath{\sigma}}$_{8}$ tension with the Planck15 /{\ensuremath{\Lambda}} CDM determination and implications for modified gravity theories}",
      journal = {Physical Review D},
         year = 2018,
        month = may,
       volume = {97},
       number = {10},
          eid = {103503},
        pages = {103503},
          doi = {10.1103/PhysRevD.97.103503},
archivePrefix = {arXiv},
       eprint = {1803.01337},
 primaryClass = {astro-ph.CO},
       adsurl = {https://ui.adsabs.harvard.edu/abs/2018PhRvD..97j3503K}
}

@article{cit:Bayes,
author = {Robert E. Kass and Adrian E. Raftery},
title = {Bayes Factors},
journal = {Journal of the American Statistical Association},
volume = {90},
number = {430},
pages = {773--795},
year = {1995},
publisher = {ASA Website},
doi = {10.1080/01621459.1995.10476572},
URL = { www.tandfonline.com/doi/abs/10.1080/01621459.1995.10476572},
}

@article{cit:AD,
author = {T. W. Anderson and D. A. Darling},
title = {{Asymptotic Theory of Certain "Goodness of Fit" Criteria Based on Stochastic Processes}},
volume = {23},
journal = {The Annals of Mathematical Statistics},
number = {2},
publisher = {Institute of Mathematical Statistics},
pages = {193 -- 212},
year = {1952},
doi = {10.1214/aoms/1177729437},
URL = {https://doi.org/10.1214/aoms/1177729437}
}

@ARTICLE{cit:Melia-RhBetter,
       author = {{Melia}, Fulvio},
        title = "{A comparison of the R$_{h}$ = ct and {\ensuremath{\Lambda}}CDM cosmologies using the cosmic distance duality relation}",
      journal = {Monthly Notices of the Royal Astronomical Society},
         year = 2018,
        month = dec,
       volume = {481},
       number = {4},
        pages = {4855-4862},
          doi = {10.1093/mnras/sty2596},
archivePrefix = {arXiv},
       eprint = {1804.09906},
 primaryClass = {astro-ph.CO},
       adsurl = {https://ui.adsabs.harvard.edu/abs/2018MNRAS.481.4855M}
}

@ARTICLE{cit:Melia-Shevchuk,
       author = {{Melia}, F. and {Shevchuk}, A.~S.~H.},
        title = "{The R$_{h}$=ct universe}",
      journal = {Monthly Notices of the Royal Astronomical Society},
         year = 2012,
        month = jan,
       volume = {419},
       number = {3},
        pages = {2579-2586},
          doi = {10.1111/j.1365-2966.2011.19906.x},
archivePrefix = {arXiv},
       eprint = {1109.5189},
 primaryClass = {astro-ph.CO},
       adsurl = {https://ui.adsabs.harvard.edu/abs/2012MNRAS.419.2579M}
}

@book{cit:Melia-book,
  title     = "The Cosmic Spacetime",
  author    = {{Melia}, Fulvio},
  year      = 2020,
  publisher = "CRC Press",
  address   = "Boca Raton, FL",
doi={10.1201/9781003081029}
}

@INPROCEEDINGS{cit:Hamilton,
       author = {{Hamilton}, A.~J.~S.},
        title = "{Linear Redshift Distortions: a Review}",
    booktitle = {The Evolving Universe},
         year = 1998,
       editor = {{Hamilton}, Donald},
       series = {Astrophysics and Space Science Library},
       volume = {231},
        month = jan,
        pages = {185},
          doi = {10.1007/978-94-011-4960-0_17},
archivePrefix = {arXiv},
       eprint = {astro-ph/9708102},
 primaryClass = {astro-ph},
       adsurl = {https://ui.adsabs.harvard.edu/abs/1998ASSL..231..185H}
}

@ARTICLE{cit:Riess,
       author = {{Riess}, Adam G.},
        title = "{The expansion of the Universe is faster than expected}",
      journal = {Nature Reviews Physics},
         year = 2020,
        month = jan,
       volume = {2},
       number = {1},
        pages = {10-12},
          doi = {10.1038/s42254-019-0137-0},
archivePrefix = {arXiv},
       eprint = {2001.03624},
 primaryClass = {astro-ph.CO},
       adsurl = {https://ui.adsabs.harvard.edu/abs/2020NatRP...2...10R}
}

@BOOK{cit:Padmanabhan,
  title     = "Structure Formation in the Universe",
  author    = "Padmanabhan, T",
  publisher = "Cambridge University Press",
  month     =  may,
  year      =  1993,
  address   = "Cambridge, England",
  language  = "en"
}

@BOOK{cit:Huterer-book,
  title     = "A Course in Cosmology",
  author    = {{Huterer}, Dragan},
  publisher = "Cambridge University Press",
  month     =  mar,
  year      =  2023,
  address   = "Cambridge, England",
  language  = "en",
doi={10.1017/9781009070232}
}

@ARTICLE{cit:Nesseris-Perivolaropoulos,
       author = {{Nesseris}, S. and {Perivolaropoulos}, L.},
        title = "{Testing {\ensuremath{\Lambda}}CDM with the growth function {\ensuremath{\delta}}(a): Current constraints}",
      journal = {Physical Review D},
         year = 2008,
        month = jan,
       volume = {77},
       number = {2},
          eid = {023504},
        pages = {023504},
          doi = {10.1103/PhysRevD.77.023504},
archivePrefix = {arXiv},
       eprint = {0710.1092},
 primaryClass = {astro-ph},
       adsurl = {https://ui.adsabs.harvard.edu/abs/2008PhRvD..77b3504N}
}

@ARTICLE{cit:Wang-Steinhardt,
       author = {{Wang}, Limin and {Steinhardt}, Paul J.},
        title = "{Cluster Abundance Constraints for Cosmological Models with a Time-varying, Spatially Inhomogeneous Energy Component with Negative Pressure}",
      journal = {Astrophysical Journal},
         year = 1998,
        month = dec,
       volume = {508},
       number = {2},
        pages = {483-490},
          doi = {10.1086/306436},
archivePrefix = {arXiv},
       eprint = {astro-ph/9804015},
 primaryClass = {astro-ph},
       adsurl = {https://ui.adsabs.harvard.edu/abs/1998ApJ...508..483W}
}

@ARTICLE{cit:Raffai2,
       author = {{Raffai}, Peter and {Pataki}, Adrienn and {B{\"o}ttger}, Rebeka L. and {Karsai}, Alexandra and {D{\'a}lya}, Gergely},
        title = "{Cosmic Chronometers, Pantheon+ Supernovae, and Quasars Favor Coasting Cosmologies over the Flat {\ensuremath{\Lambda}}CDM Model}",
      journal = {Astrophysical Journal},
         year = 2025,
        month = jan,
       volume = {979},
       number = {1},
          eid = {51},
        pages = {51},
          doi = {10.3847/1538-4357/ada249},
archivePrefix = {arXiv},
       eprint = {2412.15717},
 primaryClass = {astro-ph.CO},
       adsurl = {https://ui.adsabs.harvard.edu/abs/2025ApJ...979...51R}
}

@misc{cit:Turner_RJ,
      title={Cosmology with Peculiar Velocity Surveys}, 
      author={Ryan J. Turner},
      year={2025},
      eprint={2411.19484},
      archivePrefix={arXiv},
      primaryClass={astro-ph.CO},
      doi={10.48550/arXiv.2411.19484}, 
}

@misc{cit:zenodo,
  author       = {Ködmön, Dávid Attila and
                  Raffai, Peter},
  title        = {Code Repository for "Testing Coasting Cosmologies
                   with Large-Scale Structure Growth"
                  },
  month        = may,
  year         = 2025,
  publisher    = {Zenodo},
  doi          = {10.5281/zenodo.15548205},
  url          = {https://doi.org/10.5281/zenodo.15548205},
}

@MISC{cit:matlab,
    author = {{The MathWorks Inc.}},
    title = {adtest, MATLAB Version: 24.1.0.2628055 (R2024a) Update 4},
    year = {2024},
    publisher = {The MathWorks Inc.},
    address = {Natick, Massachusetts, United States},
    url = {https://www.mathworks.com/help/stats/adtest.html}
}

@article{cit:Li-correction,
    author = {Li, En-Kun and Du, Minghui and Zhou, Zhi-Huan and Zhang, Hongchao and Xu, Lixin},
    title = {Testing the effect of $H_0$ on $f\sigma_8$ tension using a Gaussian process method},
    journal = {Monthly Notices of the Royal Astronomical Society},
    volume = {501},
    number = {3},
    pages = {4452-4463},
    year = {2020},
    month = {12},
    issn = {0035-8711},
    doi = {10.1093/mnras/staa3894},
    url = {https://doi.org/10.1093/mnras/staa3894}
}

@ARTICLE{cit:Speagle,
       author = {{Speagle}, Joshua S.},
        title = "{DYNESTY: a dynamic nested sampling package for estimating Bayesian posteriors and evidences}",
      journal = {Monthly Notices of the Royal Astronomical Society},
         year = 2020,
        month = apr,
       volume = {493},
       number = {3},
        pages = {3132-3158},
          doi = {10.1093/mnras/staa278},
archivePrefix = {arXiv},
       eprint = {1904.02180},
 primaryClass = {astro-ph.IM},
       adsurl = {https://ui.adsabs.harvard.edu/abs/2020MNRAS.493.3132S}
}

@software{cit:Koposov,
  author       = {Sergey Koposov and
                  Josh Speagle and
                  Kyle Barbary and
                  Gregory Ashton and
                  Ed Bennett and
                  Johannes Buchner and
                  Carl Scheffler and
                  Ben Cook and
                  Colm Talbot and
                  James Guillochon and
                  Patricio Cubillos and
                  Andrés Asensio Ramos and
                  Matthieu Dartiailh and
                  Ilya and
                  Erik Tollerud and
                  Dustin Lang and
                  Ben Johnson and
                  jtmendel and
                  Edward Higson and
                  Thomas Vandal and
                  Tansu Daylan and
                  Ruth Angus and
                  patelR and
                  Phillip Cargile and
                  Patrick Sheehan and
                  Matt Pitkin and
                  Matthew Kirk and
                  Joel Leja and
                  joezuntz and
                  Danny Goldstein},
  title        = {joshspeagle/dynesty: v2.1.4},
  month        = jun,
  year         = 2024,
  publisher    = {Zenodo},
  version      = {v2.1.4},
  doi          = {10.5281/zenodo.12537467},
  url          = {https://doi.org/10.5281/zenodo.12537467},
}

@article{cit:Velasquez-Toribio,
   title={The growth factor parametrization versus numerical solutions in flat and non-flat dark energy models},
   volume={80},
   ISSN={1434-6052},
   url={http://dx.doi.org/10.1140/epjc/s10052-020-08785-z},
   DOI={10.1140/epjc/s10052-020-08785-z},
   number={12},
   journal={The European Physical Journal C},
   publisher={Springer Science and Business Media LLC},
   author={Velasquez-Toribio, A. M. and Fabris, Júlio C.},
   year={2020},
   month=dec }

@article{cit:Borges,
   title={Testing the growth rate in homogeneous and inhomogeneous interacting vacuum models},
   volume={2023},
   ISSN={1475-7516},
   url={http://dx.doi.org/10.1088/1475-7516/2023/06/009},
   DOI={10.1088/1475-7516/2023/06/009},
   number={06},
   journal={Journal of Cosmology and Astroparticle Physics},
   publisher={IOP Publishing},
   author={Borges, H.A. and Pigozzo, C. and Hepp, P. and Baraúna, L.O. and Benetti, M.},
   year={2023},
   month=jun, pages={009} }

@article{cit:Arjona,
   title={Cosmological constraints on nonadiabatic dark energy perturbations},
   volume={102},
   ISSN={2470-0029},
   url={http://dx.doi.org/10.1103/PhysRevD.102.103526},
   DOI={10.1103/physrevd.102.103526},
   number={10},
   journal={Physical Review D},
   publisher={American Physical Society (APS)},
   author={Arjona, Rubén and García-Bellido, Juan and Nesseris, Savvas},
   year={2020},
   month=nov }

@article{cit:Vehtari-Gelman-Gabry,
  author    = {Vehtari, Aki and Gelman, Andrew and Gabry, Jonah},
  title     = {Practical Bayesian model evaluation using leave-one-out cross-validation and {WAIC}},
  journal   = {Statistics and Computing},
  volume    = {27},
  number    = {5},
  pages     = {1413--1432},
  year      = {2017},
  doi       = {10.1007/s11222-016-9696-4},
  url       = {https://doi.org/10.1007/s11222-016-9696-4}
}

@article{cit:Gelman-Meng-Stern,
  author    = {Gelman, Andrew and Meng, Xiao-Li and Stern, Hal},
  title     = {Posterior predictive assessment of model fitness via realized discrepancies},
  journal   = {Statistica Sinica},
  volume    = {6},
  number    = {4},
  pages     = {733--760},
  year      = {1996},
  url       = {https://www.jstor.org/stable/24306036}
}

@ARTICLE{cit:cosmoverse,
       author = {{Di Valentino}, Eleonora and {Said}, Jackson Levi and {Riess}, Adam and {Pollo}, Agnieszka and {Poulin}, Vivian and {G{\'o}mez-Valent}, Adri{\`a} and {Weltman}, Amanda and {Palmese}, Antonella and {Huang}, Caroline D. and {van de Bruck}, Carsten and {Saraf}, Chandra Shekhar and {Kuo}, Cheng-Yu and {Uhlemann}, Cora and {Grand{\'o}n}, Daniela and {Paz}, Dante and {Eckert}, Dominique and {Teixeira}, Elsa M. and {Saridakis}, Emmanuel N. and {Colg{\'a}in}, Eoin {\'O}. and {Beutler}, Florian and {Niedermann}, Florian and {Bajardi}, Francesco and {Barenboim}, Gabriela and {Gubitosi}, Giulia and {Musella}, Ilaria and {Banik}, Indranil and {Szapudi}, Istvan and {Singal}, Jack and {Cases}, Jaume Haro and {Chluba}, Jens and {Torrado}, Jes{\'u}s and {Mifsud}, Jurgen and {Jedamzik}, Karsten and {Said}, Khaled and {Dialektopoulos}, Konstantinos and {Herold}, Laura and {Perivolaropoulos}, Leandros and {Zu}, Lei and {Galbany}, Llu{\'\i}s and {Breuval}, Louise and {Visinelli}, Luca and {Escamilla}, Luis A. and {Anchordoqui}, Luis A. and {Sheikh-Jabbari}, M.~M. and {Lembo}, Margherita and {Dainotti}, Maria Giovanna and {Vincenzi}, Maria and {Asgari}, Marika and {Gerbino}, Martina and {Forconi}, Matteo and {Cantiello}, Michele and {Moresco}, Michele and {Benetti}, Micol and {Sch{\"o}neberg}, Nils and {Akarsu}, {\"O}zg{\"u}r and {Nunes}, Rafael C. and {Bernardo}, Reginald Christian and {Ch{\'a}vez}, Ricardo and {Anderson}, Richard I. and {Watkins}, Richard and {Capozziello}, Salvatore and {Li}, Siyang and {Vagnozzi}, Sunny and {Pan}, Supriya and {Treu}, Tommaso and {Irsic}, Vid and {Handley}, Will and {Giar{\`e}}, William and {Murakami}, Yukei and {Banihashemi}, Abdolali and {Poudou}, Ad{\`e}le and {Heavens}, Alan and {Kogut}, Alan and {Domi}, Alba and {Lenart}, Aleksander {\L}ukasz and {Melchiorri}, Alessandro and {Vadal{\`a}}, Alessandro and {Amon}, Alexandra and {Rivera}, Alexander Bonilla and {Reeves}, Alexander and {Zhuk}, Alexander and {Bonanno}, Alfio and {{\"O}vg{\"u}n}, Ali and {Pisani}, Alice and {Talebian}, Alireza and {Abebe}, Amare and {Aboubrahim}, Amin and {Gonz{\'a}lez Mor{\'a}n}, Ana Luisa and {Kov{\'a}cs}, Andr{\'a}s and {Lymperis}, Andreas and {Papatriantafyllou}, Andreas and {Liddle}, Andrew R. and {Paliathanasis}, Andronikos and {Borowiec}, Andrzej and {Yadav}, Anil Kumar and {Yadav}, Anita and {Sen}, Anjan Ananda and {William}, Anjitha John and {Davis}, Anne Christine and {Shajib}, Anowar J. and {Walters}, Anthony and {Lonappan}, Anto Idicherian and {Chudaykin}, Anton and {Capodagli}, Antonio and {da Silva}, Antonio and {De Felice}, Antonio and {Racioppi}, Antonio and {Oficial}, Araceli Soler and {Montiel}, Ariadna and {Favale}, Arianna and {Bernui}, Armando and {Velasco}, Arrianne Crystal and {Heinesen}, Asta and {Bakopoulos}, Athanasios and {Chatzistavrakidis}, Athanasios and {Khanpour}, Bahman and {Sathyaprakash}, Bangalore S. and {Zgirski}, Bartek and {L'Huillier}, Benjamin and {Famaey}, Benoit and {Jain}, Bhuvnesh and {Zhang}, Bing and {Karmakar}, Biswajit and {Dragovich}, Branko and {Thomas}, Brooks and {Correa}, Carlos and {Boiza}, Carlos G. and {Marques}, Catarina and {Escamilla-Rivera}, Celia and {Tzerefos}, Charalampos and {Zhang}, Chi and {De Leo}, Chiara and {Pfeifer}, Christian and {Lee}, Christine and {Venter}, Christo and {Gomes}, Cl{\'a}udio and {Roque De bom}, Clecio and {Moreno-Pulido}, Cristian and {Iosifidis}, Damianos and {Grin}, Dan and {Blixt}, Daniel and {Scolnic}, Dan and {Oriti}, Daniele and {Dobrycheva}, Daria and {Bettoni}, Dario and {Benisty}, David and {Fern{\'a}ndez-Arenas}, David and {Wiltshire}, David L. and {Sanchez Cid}, David and {Tamayo}, David and {Valls-Gabaud}, David and {Pedrotti}, Davide and {Wang}, Deng and {Staicova}, Denitsa and {Totolou}, Despoina and {Rubiera-Garcia}, Diego and {Milakovi{\'c}}, Dinko and {Pesce}, Dominic W. and {Sluse}, Dominique and {Borka}, Du{\v{s}}ko and {Yusofi}, Ebrahim and {Giusarma}, Elena and {Terlevich}, Elena and {Tomasetti}, Elena and {Vagenas}, Elias C. and {Fazzari}, Elisa and {Ferreira}, Elisa G.~M. and {Barakovic}, Elvis and {Dimastrogiovanni}, Emanuela and {Holm}, Emil Brinch and {Mottola}, Emil and {{\"O}z{\"u}lker}, Emre and {Specogna}, Enrico and {Brocato}, Enzo and {Jensko}, Erik and {Enriquez}, Erika Antonette and {Bhatia}, Esha and {Bresolin}, Fabio and {Avila}, Felipe and {Bouch{\`e}}, Filippo and {Bombacigno}, Flavio and {Anagnostopoulos}, Fotios K. and {Pace}, Francesco and {Sorrenti}, Francesco and {Lobo}, Francisco S.~N. and {Courbin}, Fr{\'e}d{\'e}ric and {Hansen}, Frode K. and {Sloan}, Greg and {Farrugia}, Gabriel and {Lynch}, Gabriel and {Garcia-Arroyo}, Gabriela and {Raimondo}, Gabriella and {Lambiase}, Gaetano and {Anand}, Gagandeep S. and {Poulot}, Gaspard and {Leon}, Genly and {Kouniatalis}, Gerasimos and {Nardini}, Germano and {Cs{\"o}rnyei}, G{\'e}za and {Galloni}, Giacomo},
        title = "{The CosmoVerse White Paper: Addressing observational tensions in cosmology with systematics and fundamental physics}",
      journal = {Physics of the Dark Universe},
         year = 2025,
        month = sep,
       volume = {49},
          eid = {101965},
        pages = {101965},
          doi = {10.1016/j.dark.2025.101965},
archivePrefix = {arXiv},
       eprint = {2504.01669},
 primaryClass = {astro-ph.CO},
       adsurl = {https://ui.adsabs.harvard.edu/abs/2025PDU....4901965D}
}

@INPROCEEDINGS{cit:Skilling-2004,
       author = {{Skilling}, John},
        title = "{Nested Sampling}",
    booktitle = {Bayesian Inference and Maximum Entropy Methods in Science and Engineering: 24th International Workshop on Bayesian Inference and Maximum Entropy Methods in Science and Engineering},
         year = 2004,
       editor = {{Fischer}, Rainer and {Preuss}, Roland and {Toussaint}, Udo Von},
       series = {American Institute of Physics Conference Series},
       volume = {735},
        month = nov,
    publisher = {AIP},
        pages = {395-405},
          doi = {10.1063/1.1835238},
       adsurl = {https://ui.adsabs.harvard.edu/abs/2004AIPC..735..395S}
}

@article{cit:Skilling-2006,
author = {{Skilling}, John},
title = {{Nested sampling for general Bayesian computation}},
volume = {1},
journal = {Bayesian Analysis},
number = {4},
publisher = {International Society for Bayesian Analysis},
pages = {833 -- 859},
year = {2006},
doi = {10.1214/06-BA127},
URL = {https://doi.org/10.1214/06-BA127}
}

@ARTICLE{cit:Feroz-Hobson-Bridges,
       author = {{Feroz}, F. and {Hobson}, M.~P. and {Bridges}, M.},
        title = "{MULTINEST: an efficient and robust Bayesian inference tool for cosmology and particle physics}",
      journal = {Monthly Notices of the Royal Astronomical Society},
         year = 2009,
        month = oct,
       volume = {398},
       number = {4},
        pages = {1601-1614},
          doi = {10.1111/j.1365-2966.2009.14548.x},
archivePrefix = {arXiv},
       eprint = {0809.3437},
 primaryClass = {astro-ph},
       adsurl = {https://ui.adsabs.harvard.edu/abs/2009MNRAS.398.1601F}
}

@article{cit:Deliduman,
title = {Growth index in the $\gamma \delta$CDM model},
journal = {Physics of the Dark Universe},
volume = {51},
pages = {102201},
year = {2026},
issn = {2212-6864},
doi = {https://doi.org/10.1016/j.dark.2025.102201},
url = {https://www.sciencedirect.com/science/article/pii/S2212686425003930},
author = {Cemsinan Deliduman and Furkan Şakir Dilsiz and Selinay Sude Binici}
}

@article{cit:John-Joseph-2000,
  title = {Generalized Chen-Wu type cosmological model},
  author = {John, Moncy V. and Joseph, K. Babu},
  journal = {Phys. Rev. D},
  volume = {61},
  issue = {8},
  pages = {087304},
  numpages = {4},
  year = {2000},
  month = {Mar},
  publisher = {American Physical Society},
  doi = {10.1103/PhysRevD.61.087304},
  url = {https://link.aps.org/doi/10.1103/PhysRevD.61.087304}
}

@ARTICLE{cit:Verde2024,
       author = {{Verde}, Licia and {Sch{\"o}neberg}, Nils and {Gil-Mar{\'\i}n}, H{\'e}ctor},
        title = "{A Tale of Many H $_{0}$}",
      journal = {Annual Review of Astronomy and Astrophysics,},
         year = 2024,
       volume = {62},
       number = {1},
        pages = {287-331},
          doi = {10.1146/annurev-astro-052622-033813},
archivePrefix = {arXiv},
       eprint = {2311.13305},
 primaryClass = {astro-ph.CO},
       adsurl = {https://ui.adsabs.harvard.edu/abs/2024ARA&A..62..287V}
}

@article{cit:Yennapureddy2021,
title = {Structure formation and the matter power-spectrum in the Rh=ct universe},
journal = {Physics of the Dark Universe},
volume = {31},
pages = {100752},
year = {2021},
issn = {2212-6864},
doi = {https://doi.org/10.1016/j.dark.2020.100752},
url = {https://www.sciencedirect.com/science/article/pii/S2212686420304659},
author = {Manoj K. Yennapureddy and Fulvio Melia},
}

@article{cit:Pantos2026,
   title={Status of the $\mathit{S}_8$ tension: A 2026 review of probe discrepancies},
   volume={52},
   ISSN={2212-6864},
   url={http://dx.doi.org/10.1016/j.dark.2026.102286},
   DOI={10.1016/j.dark.2026.102286},
   journal={Physics of the Dark Universe},
   publisher={Elsevier BV},
   author={Pantos, Ioannis and Perivolaropoulos, Leandros},
   year={2026},
   month=June, pages={102286} }

@article{cit:Madhavacheril2024,
   title={The Atacama Cosmology Telescope: DR6 Gravitational Lensing Map and Cosmological Parameters},
   volume={962},
   ISSN={1538-4357},
   url={http://dx.doi.org/10.3847/1538-4357/acff5f},
   DOI={10.3847/1538-4357/acff5f},
   number={2},
   journal={The Astrophysical Journal},
   publisher={American Astronomical Society},
   author={Madhavacheril, Mathew S. and Qu, Frank J. and Sherwin, Blake D. and MacCrann, Niall and Li, Yaqiong and Abril-Cabezas, Irene and Ade, Peter A. R. and Aiola, Simone and Alford, Tommy and Amiri, Mandana and Amodeo, Stefania and An, Rui and Atkins, Zachary and Austermann, Jason E. and Battaglia, Nicholas and Battistelli, Elia Stefano and Beall, James A. and Bean, Rachel and Beringue, Benjamin and Bhandarkar, Tanay and Biermann, Emily and Bolliet, Boris and Bond, J Richard and Cai, Hongbo and Calabrese, Erminia and Calafut, Victoria and Capalbo, Valentina and Carrero, Felipe and Challinor, Anthony and Chesmore, Grace E. and Cho, Hsiao-mei and Choi, Steve K. and Clark, Susan E. and Córdova Rosado, Rodrigo and Cothard, Nicholas F. and Coughlin, Kevin and Coulton, William and Crowley, Kevin T. and Dalal, Roohi and Darwish, Omar and Devlin, Mark J. and Dicker, Simon and Doze, Peter and Duell, Cody J. and Duff, Shannon M. and Duivenvoorden, Adriaan J. and Dunkley, Jo and Dünner, Rolando and Fanfani, Valentina and Fankhanel, Max and Farren, Gerrit and Ferraro, Simone and Freundt, Rodrigo and Fuzia, Brittany and Gallardo, Patricio A. and Garrido, Xavier and Givans, Jahmour and Gluscevic, Vera and Golec, Joseph E. and Guan, Yilun and Hall, Kirsten R. and Halpern, Mark and Han, Dongwon and Harrison, Ian and Hasselfield, Matthew and Healy, Erin and Henderson, Shawn and Hensley, Brandon and Hervías-Caimapo, Carlos and Hill, J. Colin and Hilton, Gene C. and Hilton, Matt and Hincks, Adam D. and Hložek, Renée and Ho, Shuay-Pwu Patty and Huber, Zachary B. and Hubmayr, Johannes and Huffenberger, Kevin M. and Hughes, John P. and Irwin, Kent and Isopi, Giovanni and Jense, Hidde T. and Keller, Ben and Kim, Joshua and Knowles, Kenda and Koopman, Brian J. and Kosowsky, Arthur and Kramer, Darby and Kusiak, Aleksandra and La Posta, Adrien and Lague, Alex and Lakey, Victoria and Lee, Eunseong and Li, Zack and Limon, Michele and Lokken, Martine and Louis, Thibaut and Lungu, Marius and MacInnis, Amanda and Maldonado, Diego and Maldonado, Felipe and Mallaby-Kay, Maya and Marques, Gabriela A. and McMahon, Jeff and Mehta, Yogesh and Menanteau, Felipe and Moodley, Kavilan and Morris, Thomas W. and Mroczkowski, Tony and Naess, Sigurd and Namikawa, Toshiya and Nati, Federico and Newburgh, Laura and Nicola, Andrina and Niemack, Michael D. and Nolta, Michael R. and Orlowski-Scherer, John and Page, Lyman A. and Pandey, Shivam and Partridge, Bruce and Prince, Heather and Puddu, Roberto and Radiconi, Federico and Robertson, Naomi and Rojas, Felipe and Sakuma, Tai and Salatino, Maria and Schaan, Emmanuel and Schmitt, Benjamin L. and Sehgal, Neelima and Shaikh, Shabbir and Sierra, Carlos and Sievers, Jon and Sifón, Cristóbal and Simon, Sara and Sonka, Rita and Spergel, David N. and Staggs, Suzanne T. and Storer, Emilie and Switzer, Eric R. and Tampier, Niklas and Thornton, Robert and Trac, Hy and Treu, Jesse and Tucker, Carole and Ullom, Joel and Vale, Leila R. and Van Engelen, Alexander and Van Lanen, Jeff and van Marrewijk, Joshiwa and Vargas, Cristian and Vavagiakis, Eve M. and Wagoner, Kasey and Wang, Yuhan and Wenzl, Lukas and Wollack, Edward J. and Xu, Zhilei and Zago, Fernando and Zheng, Kaiwen},
   year={2024},
   month=Feb, pages={113} }

@article{cit:Wright2025,
   title={KiDS-Legacy: Cosmological constraints from cosmic shear with the complete Kilo-Degree Survey},
   volume={703},
   ISSN={1432-0746},
   url={http://dx.doi.org/10.1051/0004-6361/202554908},
   DOI={10.1051/0004-6361/202554908},
   journal={Astronomy \& Astrophysics},
   publisher={EDP Sciences},
   author={Wright, Angus H. and Stölzner, Benjamin and Asgari, Marika and Bilicki, Maciej and Giblin, Benjamin and Heymans, Catherine and Hildebrandt, Hendrik and Hoekstra, Henk and Joachimi, Benjamin and Kuijken, Konrad and Li, Shun-Sheng and Reischke, Robert and von Wietersheim-Kramsta, Maximilian and Yoon, Mijin and Burger, Pierre and Chisari, Nora Elisa and de Jong, Jelte and Dvornik, Andrej and Georgiou, Christos and Harnois-Déraps, Joachim and Jalan, Priyanka and William, Anjitha John and Joudaki, Shahab and Lesci, Giorgio Francesco and Linke, Laila and Loureiro, Arthur and Mahony, Constance and Maturi, Matteo and Miller, Lance and Moscardini, Lauro and Napolitano, Nicola R. and Porth, Lucas and Radovich, Mario and Schneider, Peter and Tröster, Tilman and Valentijn, Edwin and Wittje, Anna and Yan, Ziang and Zhang, Yun-Hao},
   year={2025},
   month=Nov, pages={A158} }

@article{cit:Abbott2026,
      title={Dark Energy Survey Year 6 Results: Cosmological Constraints from Galaxy Clustering and Weak Lensing},
      collaboration={DES},
      author={T. M. C. Abbott and M. Adamow and M. Aguena and A. Alarcon and S. S. Allam and O. Alves and A. Amon and D. Anbajagane and F. Andrade-Oliveira and S. Avila and D. Bacon and E. J. Baxter and J. Beas-Gonzalez and K. Bechtol and M. R. Becker and G. M. Bernstein and E. Bertin and J. Blazek and S. Bocquet and D. Brooks and D. Brout and H. Camacho and G. Camacho-Ciurana and R. Camilleri and G. Campailla and A. Campos and A. Carnero Rosell and M. Carrasco Kind and J. Carretero and P. Carrilho and F. J. Castander and R. Cawthon and C. Chang and A. Choi and J. M. Coloma-Nadal and M. Costanzi and M. Crocce and W. d'Assignies and L. N. da Costa and M. E. da Silva Pereira and T. M. Davis and J. De Vicente and J. DeRose and H. T. Diehl and S. Dodelson and P. Doel and C. Doux and A. Drlica-Wagner and T. F. Eifler and J. Elvin-Poole and J. Estrada and S. Everett and A. E. Evrard and J. Fang and A. Farahi and A. Ferté and B. Flaugher and P. Fosalba and J. Frieman and J. García-Bellido and M. Gatti and E. Gaztanaga and G. Giannini and P. Giles and K. Glazebrook and M. Gorsuch and D. Gruen and R. A. Gruendl and J. Gschwend and G. Gutierrez and I. Harrison and W. G. Hartley and E. Henning and K. Herner and S. R. Hinton and D. L. Hollowood and K. Honscheid and E. M. Huff and D. Huterer and B. Jain and D. J. James and M. Jarvis and N. Jeffrey and T. Jeltema and T. Kacprzak and S. Kent and A. Kovacs and E. Krause and R. Kron and K. Kuehn and O. Lahav and S. Lee and E. Legnani and C. Lidman and H. Lin and N. MacCrann and M. Manera and T. Manning and J. L. Marshall and S. Mau and J. McCullough and J. Mena-Fernández and F. Menanteau and R. Miquel and J. J. Mohr and J. Muir and J. Myles and R. C. Nichol and B. Nord and J. H. O'Donnell and R. L. C. Ogando and A. Palmese and M. Paterno and J. Peoples and W. J. Percival and D. Petravick and A. Pieres and A. A. Plazas Malagón and A. Porredon and A. Pourtsidou and J. Prat and C. Preston and M. Raveri and W. Riquelme and M. Rodriguez-Monroy and P. Rogozenski and A. K. Romer and A. Roodman and R. Rosenfeld and A. J. Ross and E. Rozo and E. S. Rykoff and S. Samuroff and C. Sánchez and E. Sanchez and D. Sanchez Cid and T. Schutt and I. Sevilla-Noarbe and E. Sheldon and N. Sherman and T. Shin and M. Smith and M. Soares-Santos and E. Suchyta and M. E. C. Swanson and M. Tabbutt and G. Tarle and D. Thomas and C. To and A. Tong and L. Toribio San Cipriano and M. A. Troxel and M. Tsedrik and D. L. Tucker and V. Vikram and A. R. Walker and N. Weaverdyck and R. H. Wechsler and D. H. Weinberg and J. Weller and V. Wetzell and A. Whyley and R. D. Wilkinson and P. Wiseman and H. -Y. Wu and M. Yamamoto and B. Yanny and B. Yin and G. Zacharegkas and Y. Zhang and J. Zuntz},
      year={2026},
      eprint={2601.14559},
      archivePrefix={arXiv},
      primaryClass={astro-ph.CO},
      url={https://arxiv.org/abs/2601.14559}, 
}
\end{document}